# Fermi gas of polar molecules in the Pauli-blocked regime

Junyu Lin[1]*, Annette N. Carroll[1], Phillip Martin[1], Calder Miller[1]†, Reuben R. W. Wang[2, 3], Kevin B. Xu[1], John L. Bohn[1], Tim de Jongh[1], Jun Ye[1]*

[1]JILA, National Institute of Standards and Technology, and Department of Physics, University of Colorado, Boulder, CO 80309, USA

[2]ITAMP, Center for Astrophysics | Harvard & Smithsonian, Cambridge, Massachusetts 02138, USA

[3]Department of Physics, Harvard University, Cambridge, Massachusetts 02138, USA

†Present address: California Institute of Technology, Pasadena, California 91125, USA

*Corresponding authors; e-mails: Junyu.Lin@colorado.edu, ye@jila.colorado.edu

**Abstract: Quantum gases of polar molecules have recently emerged as a powerful platform for exploring exotic many-body dynamics and correlated quantum behavior. To achieve the full potential of this platform, the production of deeply degenerate quantum gases of molecules in arbitrary confinement geometries is necessary. Here, we successfully evaporate fermionic KRb molecules in both 3D and quasi-2D geometries to well below their Fermi temperatures utilizing dipolar collisions. As we evaporate deeper into degeneracy in both geometries, we enter the Pauli-blocked regime with polar molecules, which we independently confirm by measuring the Pauli suppression of elastic collisions. Moreover, the Pauli suppression of collisions contributes to the limitation of our final molecular temperature to about 25% of the Fermi temperature in both geometries, particularly limiting quasi-2D evaporation where the Pauli blockade drastically reduces an otherwise large elastic to inelastic scattering ratio. This work demonstrates the production of degenerate Fermi gases of polar molecules both in a 3D harmonic trap and in mono- and bi-layer 2D configurations. Further, our work explores the fundamental limits on evaporation of molecular Fermi gases set by the Pauli-exclusion principle, which could be overcome in the future by introducing distinguishable scattering partners.**

## Introduction

Quantum gases of polar molecules are uniquely useful systems for the study of quantum many-body physics and controlled quantum chemistry, owing to their strong, tunable interactions and rich internal structure[1–5]. The long-range, anisotropic nature of these molecular interactions has enabled the exploration of generalized spin models[6–11] and the observation of dipolar interaction-induced many-body effects such as Fermi surface deformation[12] and quantum droplet formation[5,13]. The unique strength of molecular platforms lies in the large experimental tunability of the Hamiltonian governing the system where, in addition to the highly tunable interactions[8,9], the dimensionality and kinetic energy of the system can be controlled using optical lattices[14,6,2,7,10]. Two-dimensional molecular systems in mono- or bilayer geometry are particularly promising, owing to the enhanced control and tunability over the anisotropy of the dipolar interactions the additional confinement provides, and have been predicted to host a multitude of exotic quantum many-body phases[15–20] including topological superfluidity[21] and fractional Chern insulating states[22].

To fully unlock the capabilities of these platforms, deep quantum degeneracy in a range of controlled potential landscapes must be attained. For atomic quantum gases, this is routinely achieved through efficient evaporative cooling in both three-dimensional (3D)[23–25] and two-dimensional (2D) configurations[26–28]. The rich molecular interactions, however, come at the cost of inelastic (reactive) collisions which can significantly enhance losses[29–32]. Recent years have therefore seen a development of molecular shielding techniques, which apply static[2,33,34] or dynamic external fields[3,35–37] to engineer favorable elastic-to-inelastic cross-section ratios. These have led to major breakthroughs in molecular evaporation, where both microwave and static electric field shielding have allowed efficient cooling in 3D[34,36,37] and the achievement of quantum degeneracy[3–5]. Thus far, only static electric field shielding has produced quasi-2D degenerate gases of fermionic molecules, at temperatures of 0.6 $T_{\mathrm{F}}$, with $T_{\mathrm{F}}$ being the system's Fermi temperature[2]. This was achieved through confinement of the molecules in multiple layers of a one-dimensional optical lattice, yet full control over molecule number layer distribution[38] in concurrence with evaporation to degeneracy has remained elusive.

Here, we report the production of degenerate gases of polar molecules well below their Fermi temperatures in 3D and mono- and bilayer quasi-2D configurations, ultimately achieving a $T/T_{\mathrm{F}} = 0.33(2)$ in 3D and $T/T_{\mathrm{F}} = 0.24(8)$ in quasi-2D, with the latter representing an almost threefold reduction in the lowest reported reduced temperature of a quasi-2D molecular sample to date[2]. In this new regime, we additionally report the direct observation of Pauli-suppressed elastic collisions between molecules. Limits to the lowest achievable temperatures with dipolar evaporation are found to be set by both the Pauli suppression of elastic collisions[39–41], arising from the reduction of unoccupied states in the trap needed for collisional thermalization, and from one-body heating in the optical traps[42,14]. In 3D, the induced electric dipole moment of the resonantly shielded molecules is relatively small, which limits their optimal evaporation rate, and results in technical heating reducing the evaporation efficiency to a similar extent as the Pauli blockade. By contrast, in quasi-2D, the reduced dimensionality allows the molecules to be polarized to a larger dipole moment, promising an enhanced elastic collision rate and subsequently improved evaporative cooling efficiency that is not as limited by technical heating. However, the reduced dimensionality also leads to a lower density of states, causing severe Pauli suppression of elastic collisions. We confirm this competition in quasi-2D by observing that a high evaporation efficiency is eventually reduced by the large Pauli suppression of elastic collisions. Our observations indicate that the Pauli blockade has become a major obstacle for reaching more deeply degenerate gases of spin-polarized polar molecules. Nevertheless, evaporation with fermionic spin mixtures is known to overcome this limitation[39,43,44], and the current low entropies already open the opportunity for new exploration of exotic dipolar phenomena.

## Results

### Production of ultracold molecules in 3D and quasi-2D

We produce ultracold polar molecules in either 3D or quasi-2D geometries on the same experimental platform by modifying our optical lattice configurations. In all geometries, the $^{40}$K$^{87}$Rb (KRb) molecules were created from an ultracold mixture of Fermi-degenerate $^{40}$K and a Bose-Einstein condensate of $^{87}$Rb. The degenerate mixture was initially held in a crossed optical dipole trap (xODT).

To create 3D samples in the xODT, we first magneto-associated the $^{87}$Rb and $^{40}$K into the weakly bound Feshbach molecules[45]. They were then coherently transferred to their absolute ground state through stimulated Raman adiabatic passage[46]. The trap frequencies for KRb in the xODT were $(\omega_x, \omega_y, \omega_z) = 2\pi \times (30{,}200{,}30)$ Hz. The unpaired $^{87}$Rb and $^{40}$K were then removed from the trap by applying resonant optical pulses, after which molecular evaporation was initiated.

To generate our samples in a quasi-2D geometry, further compression of the atomic mixture along the vertical axis is required before initiating molecule production. To this end, we transferred the atomic mixture from the xODT into an accordion lattice (AL) with tunable spacing. The spatial period of the AL was controlled by the intersection angle between the two lattice beams (Fig. 1a), and it was initially set to 8 μm such that the Rb atoms were loaded into a single AL layer. The spacing of the AL was then reduced to increase the vertical confinement of the atomic mixture, compressing its spatial distribution. The mixture was then loaded into a vertical lattice (VL) with a fixed spatial period of 540 nm, which serves to confine the particles in a quasi-2D geometry. Depending on the final spacing and laser power of the AL, the mixture can be loaded into a controllable number of VL layers, ranging from 5 to 1 (Fig. 1b). After loading into the VL, we associated the mixture into KRb molecules. The trap frequencies for KRb in the VL were set to $(\omega_x,\ \omega_y,\ \omega_z) = 2\pi \times (30, 15000, 30)$ Hz. We typically obtained a molecular sample with temperature $T \approx 200$ nK in the VL before evaporation. The molecules are confined as a quasi-2D gas given $k_B T \ll \hbar\omega_y$ where $k_B$ is the Boltzmann constant and $\hbar$ the reduced Planck constant.

### Dipolar collisions in different dimensionalities

Inelastic collisions causing molecular loss can be suppressed by generating repulsive dipole-dipole interactions with static electric fields[2,33,34], preventing molecules from reaching small separations where chemical reactions occur[29–32]. In quasi-2D, an electric field of 6.5 kV/cm perpendicular to the confining plane was applied to induce an electric dipole moment $d$ of 0.25 Debye for KRb molecules in the $|\tilde{N}=0, m_{\tilde{N}}=0\rangle$ state. Here, $\tilde{N}$ is the electric-field dressed rotational quantum number and $m_{\tilde{N}}$ is the projection onto the quantization axis set by the electric field. A repulsive dipole-dipole potential was thus generated as the collisions were constrained to be predominantly side-by-side, and the chosen electric field provided the optimal evaporation efficiency[2]. For 3D geometries, the dipole-dipole interactions were engineered to be repulsive for all collisions regardless of scattering direction[34], utilizing $|1,0\rangle$ molecules with $d$ = 0.1 Debye at 12.72 kV/cm, detuned from an electric-field induced two-body resonance.

The dipole-dipole interactions not only suppress the inelastic collisions but also provide a large elastic scattering cross section for collisions in the ultracold regime. For two identical fermions interacting through van der Waals interactions, elastic scattering cross section vanishes for collisional energies approaching zero[47]. By contrast, for dipole-dipole interactions, the cross section remains finite[48]. Nevertheless, elastic scattering rates are still expected to decrease when approaching deep degeneracy due to Pauli suppression. For fermionic molecules in a harmonic trap, the Pauli exclusion principle dictates that no two molecules may occupy the same eigenstate. As illustrated in Fig. 1c, elastic scattering requires unoccupied final states, which is easily satisfied for $T > T_{\mathrm{F}}$, resulting in efficient evaporative cooling in 3D[34] and quasi-2D[2]. As the gas reaches deeper degeneracy, the number of unoccupied states under the Fermi energy $k_{\mathrm{B}}T_{\mathrm{F}}$ decreases, thus reducing the state-changing elastic collision rate by a factor of $\eta$[39–41]. While this is expected to enhance collective spin dynamics by reducing collision-induced dephasing of molecules in coherent superpositions[49], this Pauli suppression of elastic collisions poses a challenge for reaching deep degeneracy through evaporative cooling of spin-polarized samples. The magnitude of this effect differs for the distinct geometries used in our work, as the density of states $g_{\mathrm{HO}}$ of the confining harmonic oscillator trap is dimension dependent (Fig. 1d). Indeed, for systems with the same $T/T_{\mathrm{F}}$ , the fraction of unoccupied states is smaller in quasi-2D compared to 3D configurations. This leads to a larger suppression of elastic scattering in quasi-2D compared to 3D (Fig. 1e). As a result, a larger reduction of the cooling efficiency is expected for quasi-2D evaporation across the full range of experimentally achievable temperatures. Nevertheless, due to the larger elastic-to-inelastic collisional ratio enabled by the larger induced dipole moment in quasi-2D than in 3D, evaporative cooling remains effective, allowing us to produce the most deeply degenerate gas in quasi-2D rather than in 3D.

## Evaporative cooling of KRb molecules in 3D

Evaporative cooling in 3D was initiated with a molecule number of $N = 5.4(7) \times 10^4$, at a temperature $T = 177(14)$ nK corresponding to $0.82(6)T_{\mathrm{F}}$, with 3D Fermi temperature $T_{\mathrm{F}} = \hbar\bar{\omega}(6N)^{\frac{1}{3}}/k_{\mathrm{B}}$. We took advantage of the anisotropy of dipole-dipole interactions by tilting the electric field by an angle of $30°$ compared to the direction of gravity, to reduce the number of collisions required for rethermalization by a factor of two[34]. This increases the thermalization rate and enhances the evaporative cooling efficiency. The evaporation was initiated by lowering the optical trap depth, with the entire evaporation sequence taking 3 s (see Supplementary Section I). At each step of the evaporation sequence, we measured the number and temperature of the KRb gas by fitting density profiles obtained through absorption imaging after time-of-flight (TOF) with a Fermi-Dirac distribution (see Supplementary Section E). We additionally performed density fluctuation measurements[50–52] (see Supplementary Section F), which display an increasingly sub-Poissonian nature with decreasing $T/T_{\mathrm{F}}$, serving as an independent probe of Fermi degeneracy, and showing good agreement with temperature obtained from fits to a Fermi-Dirac distribution.

The cooling trajectory is shown in Fig. 2a, b with experimental data illustrated as blue circles and triangles representing two independent methods of thermometry. As $N$ decreased from $5.4(7) \times 10^4$ to $1.4(2) \times 10^4$, we observed efficient evaporation with temperatures decreasing from 0.82(6) $T_{\mathrm{F}}$ to 0.33(2) $T_{\mathrm{F}}$. Further evaporation cooled the sample down to 22(3) nK, but the reduced temperature saturated at $T/T_{\mathrm{F}} \approx 0.3$. To understand this saturation in the evaporative cooling trajectory, we performed Monte Carlo simulations[53] (see Supplementary Section K) that fully incorporate elastic and inelastic molecular collisions[34]. We additionally included an independently measured one-body heating rate of 2.0(2) nK/s in the xODT (see Methods). Results of these simulations are shown in Fig. 2a, b as blue shaded areas and are in good agreement with the measured evaporation trajectory.

To investigate the effects of the Pauli suppression of collisions on the evaporation, we additionally performed a Monte-Carlo simulation which excludes this effect (see Supplementary Section K). The simulated cooling trajectory is shown as yellow shaded areas in Fig. 2a, b. Comparing the simulations, $T$ is expected to reach 0.24(2) $T_{\mathrm{F}}$ instead of 0.28(1) $T_{\mathrm{F}}$ at the end of the evaporation if there is no Pauli suppression of elastic scattering. A quantitative comparison of the evaporative cooling efficiency $\gamma_{\mathrm{evap}} = \frac{\mathrm{d}\log T/T_{\mathrm{F}}}{\mathrm{d}\log N}$ for both measurement and simulations at different stages of evaporation is shown in Fig. 2c. At early stages of the evaporation ($N > 1.6 \times 10^4$), no substantial differences are observed between the Monte Carlo simulations including and excluding the effects of Pauli suppression. As the evaporation continues, the simulation that considers Pauli suppression is still close to the measured $\gamma_{\mathrm{evap}}$, while the simulation that ignores the effects of Pauli suppression starts to deviate.

While these simulation results do indicate Pauli suppression reduces the cooling efficiency in 3D, they also suggest that another mechanism must be limiting the cooling. An additional simulation in which we excluded the one-body heating rate (see Methods) shows that this heating rate indeed causes a reduction of $\gamma_{\mathrm{evap}}$ which is on par with the reduction caused by Pauli suppression. We therefore conclude that Pauli suppression and one-body heating reduce the cooling efficiency at a similar magnitude in 3D during the later stages of the evaporation trajectory.

The Pauli suppression of elastic collisions, which was already indicated by the reduction of the evaporative cooling efficiency, was also directly measured from the ensemble-averaged elastic scattering cross section $\sigma_{\mathrm{e}}$ with a cross-dimensional thermalization experiment[2,3,34,36,37]. A temperature imbalance for the KRb cloud between the $y$- and the $x$-, $z$- directions was initiated by compressing the trap along the y-direction. Elastic collisions redistribute the kinetic energy, decreasing the temperature difference between the different directions. From the rate at which the sample directions thermalize, governed by a set of coupled differential equations, we extract $\sigma_{\mathrm{e}}$ (see Supplementary Section G). By systematically varying the sample temperature, which was controlled by modifying the amount of trap compression, we obtained $\sigma_{\mathrm{e}}$ at a range of temperatures. The suppression of $\sigma_{\mathrm{e}}$ with decreasing temperature is summarized in Fig. 2d. As Fig. 2d shows, $\sigma_{\mathrm{e}}$ begins to decrease from the universal scattering prediction when $T < T_{\mathrm{F}}$ because of Pauli suppression of elastic collisions. The blue solid and gray dashed curves are from theoretical predictions with and without considering Pauli blockade, respectively. The prediction of $\sigma_e$ is in good agreement with our measurements.

One major limitation to our evaporation efficiency in 3D is the limited ratio between the elastic scattering cross section $\sigma_{\mathrm{e}}$ and the inelastic scattering cross section $\sigma_{\mathrm{i}}$ throughout the evaporation sequence. We also experimentally extracted $\sigma_{\mathrm{i}}$ by fitting the measured molecule number evolution with a two-body loss model (see Supplementary Section H). We see excellent agreement between the measured inelastic cross sections and the theoretically predicted values as shown in Fig. 2d. The ratio $\sigma_{\mathrm{e}}/\sigma_{\mathrm{i}}$ is around 30 for $T$ =200 nK at the beginning of the evaporation. While this ratio changes with temperature due to the different temperature-dependencies of $\sigma_{\mathrm{e}}$ and $\sigma_{\mathrm{i}}$, it never exceeds 60 and therefore limits our cooling rate. At the end of the evaporation, the evaporative cooling rate is almost equal to the one-body heating rate, causing a saturation of $T/T_{\mathrm{F}} \approx 0.3$ in 3D.

**Evaporative cooling of KRb molecules in quasi-2D**

Compared to 3D, we expect more efficient evaporation for molecules in a quasi-2D confinement because of the possibility to induce a larger dipole moment, resulting in an order-of-magnitude more favorable ratio $\sigma_{\mathrm{e}}/\sigma_{\mathrm{i}}$ of 300 at the beginning of the evaporation. This enhancement stems from the fact that in 3D the dipole moment of the resonantly shielded molecules in the $|1, 0\rangle$ state is limited to 0.1 Debye, whereas in quasi-2D, we can use the $|0, 0\rangle$ state with a larger induced dipole moment of 0.25 Debye[2]. However, Fermi statistics suppress elastic collisions in 2D more significantly than in 3D (Fig 1E) and is therefore expected to more strongly limit the 2D evaporation efficiency. By comparing the evaporation trajectories in quasi-2D to those in 3D, we can further explore the interplay of dipolar interactions and quantum statistics.

To investigate the evaporation efficiency in quasi-2D, we first prepared molecular samples in a bilayer configuration in a deep VL. Molecules were nearly equally distributed between the two layers (Supplementary Section B), with $6.8(1) \times 10^3$ molecules in each layer. This corresponds to each layer having an initial $T/T_\mathrm{F} = 1.21(1)$, where $T_\mathrm{F} = \hbar\bar{\omega}\sqrt{2N}/k_\mathrm{B}$ in 2D. By applying an anti-trapping electric field curvature along the $x$-direction[2], the overall trap depth was lowered and evaporative cooling was initiated, with the entire evaporation sequence taking 1.7 s (Supplementary Section I). The number and temperature at each step of the evaporation were extracted from fitting Fermi-Dirac distributions to absorption images of KRb Feshbach molecules after TOF.

Fig. 3a and 3b show the evaporative cooling trajectory for molecules in a quasi-2D bilayer configuration. The temperature of the molecules reached 0.24(8) $T_\mathrm{F}$ at the final step of the evaporation, substantially lower than the previous quasi-2D record of 0.6(2) $T_\mathrm{F}$ (Ref. 2). This evaporation to deeper degeneracy is enabled by the AL, which compressed the atomic mixture into fewer layers of the VL, leading to a larger number of molecules per VL layer at the start of evaporation. Comparing quasi-2D and 3D, we obtain similar final values of $T/T_\mathrm{F}$. But as the density of states scales differently in quasi-2D than in 3D (Fig. 1d), the entropy per particle in the center of the trap in quasi-2D at $T/T_\mathrm{F}$= 0.24 is $s_\mathrm{2D} = 0.84\ k_\mathrm{B}$, which is only 68% of the entropy per particle in 3D for the same $T/T_\mathrm{F}$. This indicates we have achieved a deeper level of degeneracy of KRb molecules in quasi-2D than in 3D.

As shown by the plateauing values of $T/T_\mathrm{F}$ over the last few steps of evaporation in Fig. 3b, the evaporative cooling efficiency for the quasi-2D sample was suppressed as we evaporated deeper into degeneracy, as was the case in 3D. As for the 3D case, Monte Carlo simulations were applied to the quasi-2D configuration to understand the reduction in efficiency. The green curves in Fig. 3a, b show the simulated trajectory which included both the Pauli suppression of elastic collisions and the experimentally measured one-body heating rate of 2.0(4) nK/s for molecules in the VL. The simulation excellently captures the measured evaporation trajectory, including the observation of the $T/T_\mathrm{F}$ saturation as the system reaches deeper degeneracy.

The effect from Pauli suppression of elastic scattering is verified by performing Monte Carlo simulations without the Pauli-suppressed elastic cross sections, shown as yellow curves in Fig. 3a, b, which deviate from the experimental measurements at low values of $T/T_\mathrm{F}$. The simulation predicts $T$ could reach 0.12 $T_\mathrm{F}$ at the last stage of evaporation with ~150 molecules left in each layer if elastic collisions are not modified by Fermi statistics, which is substantially lower than the experimentally measured $T/T_\mathrm{F}$ = 0.24(8). Moreover, unlike the experimental data, the yellow curves that ignore Pauli suppression do not show signs of saturation of $T/T_\mathrm{F}$ at lower temperatures.

To quantitatively compare how this Pauli suppression of elastic collision reduces the cooling efficiency, as in 3D, we show $\gamma_\mathrm{evap}$, the evaporation efficiency, at different stages of evaporation in quasi-2D (Fig. 3c). The simulation that includes Pauli suppression (green curve) is in good agreement with the measurement, showing a reduction of $\gamma_{\mathrm{evap}}$ as the molecules are evaporated to deeper degeneracy. The simulation that ignores Pauli suppression of elastic scattering (yellow curve) shows the cooling efficiency being maintained at high values of $\gamma_{\mathrm{evap}}$ even at the later stages of the evaporation sequence, a factor of four higher than when Pauli suppression is considered. This shows that Pauli suppression plays a significant role in dipolar evaporation in quasi-2D, whereas the one-body heating is less important, which is further confirmed by an additional simulation where excluding the one-body heating rate shows a similar trajectory (see Methods).

We can further confirm the Pauli suppression of elastic collisions with a direct measurement of the cross-dimensional thermalization rate. Unlike in 3D[39–41], the Pauli-suppression of elastic collisions in 2D has not yet been directly observed. As in the 3D case, we perform cross-dimensional thermalization measurements and extract $\sigma_{\mathrm{e}}$ from the measured thermalization rate. The measured temperature-dependent $\sigma_{\mathrm{e}}$, displayed in Fig. 3d, shows $\sigma_{\mathrm{e}}$ on the order of $10^{-5}$ cm for $T > T_{\mathrm{F}}$. As the temperature decreases to values below $T_{\mathrm{F}}$, we observe a clear reduction of $\sigma_{\mathrm{e}}$ that is consistent with Pauli suppression of elastic collisions. Theoretically ensemble-averaged predictions of $\sigma_{\mathrm{e}}$, which include Pauli suppression, are shown as the blue solid curve in Fig. 3d and capture the trends of our measurements well. For comparison, the value of $\sigma_{\mathrm{e}}$ for a classical gas, which does not include effects of Pauli suppression, is shown as the gray dashed curve in Fig. 3d. The classical expectation agrees well with the experimental data for temperatures greater than $T_{\mathrm{F}}$ but clearly overestimates the experimentally measured values of $\sigma_{\mathrm{e}}$ for $T < T_{\mathrm{F}}$ when quantum statistics become important.

The predicted elastic-to-inelastic cross-section ratio $\sigma_{\mathrm{e}}/\sigma_{\mathrm{i}}$ is an order of magnitude more favorable in quasi-2D compared to 3D. Having obtained the ensemble-averaged value of $\sigma_e$ we confirmed this favorable ratio experimentally by measuring the inelastic cross-section $\sigma_{\mathrm{i}}$, obtained through a fit of the number evolution to a two-body loss model (see Supplementary Section H), shown in Fig. 3d. The measured $\sigma_{\mathrm{i}}$ matches the theoretical prediction well and the ratio $\sigma_{\mathrm{e}}/\sigma_{\mathrm{i}}$ reaches 300 for $T = 200$ nK, the initial temperature at which evaporation is initiated. While this ratio remains high throughout the evaporation trajectory, the decrease of $\sigma_{\mathrm{e}}$ due to Pauli suppression is seen to limit our evaporation, leading to the observed saturation of $T/T_{\mathrm{F}}$.

**Mono- and Bi-layer Quasi-2D Samples of Degenerate Polar Molecules**

We further demonstrate our full control over the molecular layer distribution through creation of degenerate gases of ultracold KRb in either a mono- or a bilayer geometry. We experimentally switched between these two geometries through fine control of the relative position between the AL and VL, as well as through additional compression of the atoms in the AL with increased laser intensity. Because resolving the *in-situ* 540 nm interlayer separation lies beyond our imaging capabilities, layer distribution measurements required additional magnification. We achieved this through a matter-wave magnification technique based on the method introduced in Ref. 54, but followed by an additional TOF, similar to previously introduced matter-wave focusing techniques[55–57,2]. This technique enlarges the separation between the VL layers while minimizing the spread of the intralayer density distribution (see Supplementary Section A). Fig. 4a shows the resulting VL layer distribution of the molecules for bilayer (upper panel) and monolayer (lower panel) configurations.

We performed evaporative cooling of the monolayer gas, using the sequence described for our bilayer samples. Fig. 4b shows the integrated density profile of the resulting cloud, as well as a fitted Fermi-Dirac distribution (purple solid curve). From the fit, we obtained a reduced temperature of $T/T_{\mathrm{F}} = 0.44(14)$, demonstrating a degenerate monolayer quasi-2D gas of polar molecules. We attribute the slightly higher final $T/T_{\mathrm{F}}$ with ~200 molecules in the monolayer compared to the bilayer system to a higher initial $T/T_{\mathrm{F}}$ arising from additional heating during the monolayer preparation (see Supplementary Section C). For additional evidence of degeneracy, Fig. 4b shows a Gaussian profile fit only to the low-density thermal region at the edges of the cloud (gray dashed curve). This captures the absolute temperature of the cloud[58] – which is in good agreement with the temperature obtained from our fits to a Fermi-Dirac distribution – but ignores the effects of Pauli repulsion within the fermionic gas. The overestimated central density the Gaussian fit displays serves as an additional indication of Fermi degeneracy[58,1,2].

## Discussion

In this work, we have brought polar molecules into a new regime where collisions are manifestly dictated by Fermi statistics by producing degenerate Fermi gases of polar molecules in 3D and quasi-2D, reaching temperatures of about 0.25 $T_{\mathrm{F}}$ through dipolar evaporation. In both dimensionalities, we directly observed the Pauli suppression of elastic collisions between ultracold molecules. Our work suggests that we are reaching the limit set by the Pauli blockade for the evaporation efficiency of spin-polarized samples of fermionic polar molecules. We found that this limit is particularly pronounced in quasi-2D which otherwise has very favorable scattering conditions for evaporation compared to 3D. Even deeper degeneracy may be reached through enhanced thermalization rates, which can be provided by larger dipole moments or hyperfine spin mixtures[43,39,44]. The Monte Carlo simulations verified in this work are a useful tool to consider for other molecular species and trapping conditions. We note another viable approach is through entropy redistribution[59–61].

We have furthermore demonstrated controlled production of degenerate mono- and bilayer quasi-2D gases of polar molecules. With the lowest temperatures achieved, adiabatic loading into a single layer of a 3D optical lattice is expected to generate peak filling factors of 90%. At such low entropies, we expect the Pauli suppression of elastic collisions to also suppress collisions between molecules in coherent superpositions[10], opening opportunities for future studies of collective spin dynamics including spin-squeezed molecules for quantum metrology[49,62,63]. This work also paves the way for the exploration of interlayer $s$-wave[16] and intralayer $p$-wave superfluidity[21], frustrated magnetism[64] and novel quantum phases like fractional Chern insulators[22].

## Methods

### Improved atomic mixture preparation

The general experimental procedures for creating ultracold KRb molecules are described in Ref. 1,2,8,34. For this work, in addition to implementing a tunable AL for the quasi-2D sample preparations (see next section), we modified our atomic mixture preparation to improve initial conditions of our samples in all geometries. The first change was switching to a larger magnetic field for atomic evaporation, and lattice loading and compression. While previous experiments performed final atomic mixture evaporation at 556 G, about 10 G above the interspecies Feshbach resonance, we found that the interspecies scattering length at this field enhanced three-body Rb-Rb-K loss[45]. To suppress this, we instead used a higher magnetic field of 613 G, leading to a smaller interspecies scattering length, nearly doubling the number of Bose condensed Rb at the end of the AL spacing compression.

We additionally identified that a residual 1 G/cm magnetic field gradient in the $x$-direction was causing unwanted atomic loss and reduced K and Rb thermalization efficiency due to spatial displacement of the clouds, arising from the difference in the magnetic moment between the two species. To compensate for this gradient, we installed an additional optical beam that is displaced from the center of the xODT. This beam is then turned off immediately after STIRAP, as the ground-state KRb molecules have a negligible magnetic moment.

### Optical design of the accordion lattice

In order to prepare atoms in only a few layers of the fixed-spacing VL, we implemented an AL with tunable spacing and phase control. The optical design, inspired by designs used in quantum gas microscope experiments[65], is shown in Fig. 5, with additional details given in Ref. 66. The lattice is formed by focusing two spatially-separated beams, split from a single beam that passes through a dove prism pair with a 50:50 beamsplitter coating at its interface, so that they interfere at the position of atoms in the xODT.

The lattice constant $a = \frac{\lambda}{2\sin\frac{\alpha}{2}}$, with $\lambda = 1064$ nm being the laser wavelength and $\alpha$ the angle at which the beams interfere, can be tuned by changing the distance between the two arms. We can control the spacing of the beams by rotating a mirror on a galvanometer (galvo) mount before the beam hits the dove prism beamsplitter. Given practical constraints of our design, we can tune the lattice constant between about 30 μm to about 2.2 μm. As described in the main text, we find that we can load all the Rb into one layer of the AL from the xODT at a spacing of 8 μm and subsequently load only one or two VL layers if we compress the AL to 3.2 μm. Care was taken to compress the spacing as adiabatically as possible to minimize heating during the experimental sequence.

To maximize initial loading efficiency, reduce heating during compression, and control the final VL layer distribution, we find it is also important to have dynamic control of the phase of the AL which sets the position of the central layer. We achieve this by adding a phase plate, which is a piece of antireflective-coated glass, mounted on a galvo to the path of the upper arm but not the lower arm. By rotating the phase plate more or less into the optical path of the upper beam, the light experiences different path lengths, thereby adjusting the relative phase between the upper and lower beams.

**One-body heating rate measurements**

To measure the one-body heating rate induced by noise of the trapping laser, the temperature of the molecules was measured after different hold times in the xODT or in the VL, as shown in Fig. 6a and b, respectively. The heating rate is extracted by a linear fit of the temperature evolution, giving 2.0(2) nK/s for a 3D gas in the xODT and 2.0(4) nK/s for a quasi-2D gas in the VL. For the 3D heating rate measurement, we prepared a low-density sample to reduce the two-body loss during the hold time, making the heating contribution from anti-evaporation negligible.

**Role of one-body heating versus Pauli suppression on evaporation trajectories**

In this section, we show multiple Monte Carlo simulations with and without considering the effects of Pauli suppression of elastic collisions and of the one-body heating.

For molecules in the 3D trap, measured evaporative cooling trajectories and cooling efficiencies are shown in Fig. 7a and b, respectively, which are reproduced from Fig. 2b and c of the main text. As in Fig. 2b and c of the main text, the blue and yellow shaded areas in Fig. 7a are simulations with and without considering the Pauli suppression of elastic collisions, respectively. The one-body heating rate is set to 2 nK/s for both simulations. The blue and yellow curves in Fig. 7b are from the same simulations. As discussed in the main text, by comparing the results from these two simulations, we can confirm that the Pauli suppression of elastic collisions reduced the cooling efficiency in 3D. The purple shaded area in Fig. 7a is from a simulation that does not include one-body heating but does include the Pauli blockade. It shows a lower saturated temperature compared to the blue shaded area (with Pauli suppression and heating). In Fig. 7b, the purple curve (with Pauli suppression but without heating) gives a higher cooling efficiency than the blue curve (with Pauli suppression and heating) at the end of the evaporation, showing that one-body heating reduces the cooling efficiency as well.

By comparing the purple shaded area and curve (with Pauli suppression but without heating) with the yellow shaded area and curve (without Pauli suppression but with heating) in Fig. 7, we find that they show nearly identical cooling trajectory and similar cooling efficiency throughout the evaporation process. This suggests that one-body heating reduces the cooling efficiency a similar amount as Pauli suppression reduces the efficiency in 3D. A simulation considering neither the Pauli suppression nor the one-body heating is shown as the red shaded area in Fig. 7a and red curve in Fig. 7b. From this simulation, we clearly see that without considering these two effects, the reduced temperature $T/T_{\mathrm{F}}$ would not saturate near the end of evaporation. Without these effects, the cooling efficiency would maintain a value of around 0.5 throughout the evaporation sequence.

For KRb in quasi-2D, the cooling trajectory and the cooling efficiency from the measurement and simulations are shown in Fig. 8a and b, with the data reproduced from Fig. 3b and c of the main text. As in Fig. 3b and c of the main text, the green shaded area and the curve in Fig. 8a and b, respectively, are from a simulation that considers both Pauli suppression and one-body heating of 2 nK/s. This simulation matches our experimental data very well. The yellow shaded area and the curve in Fig. 8a and b, respectively, are from a simulation that does not consider Pauli suppression but does include one-body heating. As discussed in the main text, without considering the Pauli suppression, the final reduced temperature of evaporative cooling is not saturated, and the cooling efficiency is maintained to a high level of around 0.8, suggesting that Pauli suppression limits our achieved evaporation efficiency.

A simulation that includes Pauli suppression of collisions but excludes one-body heating is shown as the purple shaded area in Fig. 8a and purple curve is Fig. 8b. It shows a lower temperature after the evaporation and better cooling efficiency compared to the measurement. However, by comparing the purple curve in Fig. 8b to the yellow curve (without considering Pauli suppression, with considering one-body heating), we see that one-body heating provides a smaller reduction to the cooling efficiency than Pauli suppression does. This shows that Pauli suppression plays a more significant role than one-body heating in the evaporation in quasi-2D. A simulation considering neither Pauli suppression nor the one-body heating is shown as the red shaded area in Fig. 8a and the red curve in Fig. 8b. It displays substantially lower final temperatures after evaporation and gives a higher cooling efficiency throughout the cooling process compared to the experimentally obtained values.

Our simulations in 2D and 3D hint that improvements to the evaporative cooling of KRb molecules can be achieved in the future either by preparing a mixture (e.g. of hyperfine states) of polar molecules to reduce the effect from Pauli suppression or by minimizing noise from trapping lasers to reduce one-body heating.

**Data availability:**

Data presented in the figures are available on Zenodo: https://doi.org/10.5281/zenodo.23357231.

**Code availability:**

Code is available from the corresponding authors upon request.

**Acknowledgements:**
We thank J. S. Higgins for early experimental contributions and N. Song and C. Nishanova for experimental assistance. We thank A. M. Rey for discussion and L. Homeier and L. Hillberry for reading the manuscript and providing comments.

**Funding Statement:**
This work is supported by the US Department of Energy, Office of Science, National Quantum Information Science Research Centers, Quantum Systems Accelerator. Additional support is from the NSF Quantum Leap Challenge Institute Award 2552714; ARO and AFOSR MURIs; JILA Physics Frontier Center grant PHY-2317149; and the National Institute of Standards and Technology. A.N.C. acknowledges support from the National Science Foundation Graduate Research Fellowship under grant DGE 2040434. C.M. acknowledges support from the US Department of Defense through the National Defense Science and Engineering Graduate Fellowship. R.R.W.W. acknowledges support from the National Science Foundation through a grant from ITAMP at Harvard University. K.B.X. and J.L.B. acknowledge support from the National Science Foundation under grant PHY-2408636.

**Author contributions:**
The project was conceived by J.L, A.N.C., P.M., C.M., T.d.J., and J.Y., with J.L, A.N.C., P.M., T.d.J., and J.Y. conducting the experiments and analyzing the data. R.R.W.W., K.B.X., and J.L.B. developed the theory models. All authors contributed to interpreting results and writing the manuscript.

**Competing interests:** The authors declare that they have no competing interests.

**Additional information:**

Supplementary Information is available for this paper.

Correspondence and requests for materials should be addressed to Junyu Lin (Junyu.Lin@colorado.edu) and Jun Ye (ye@jila.colorado.edu).

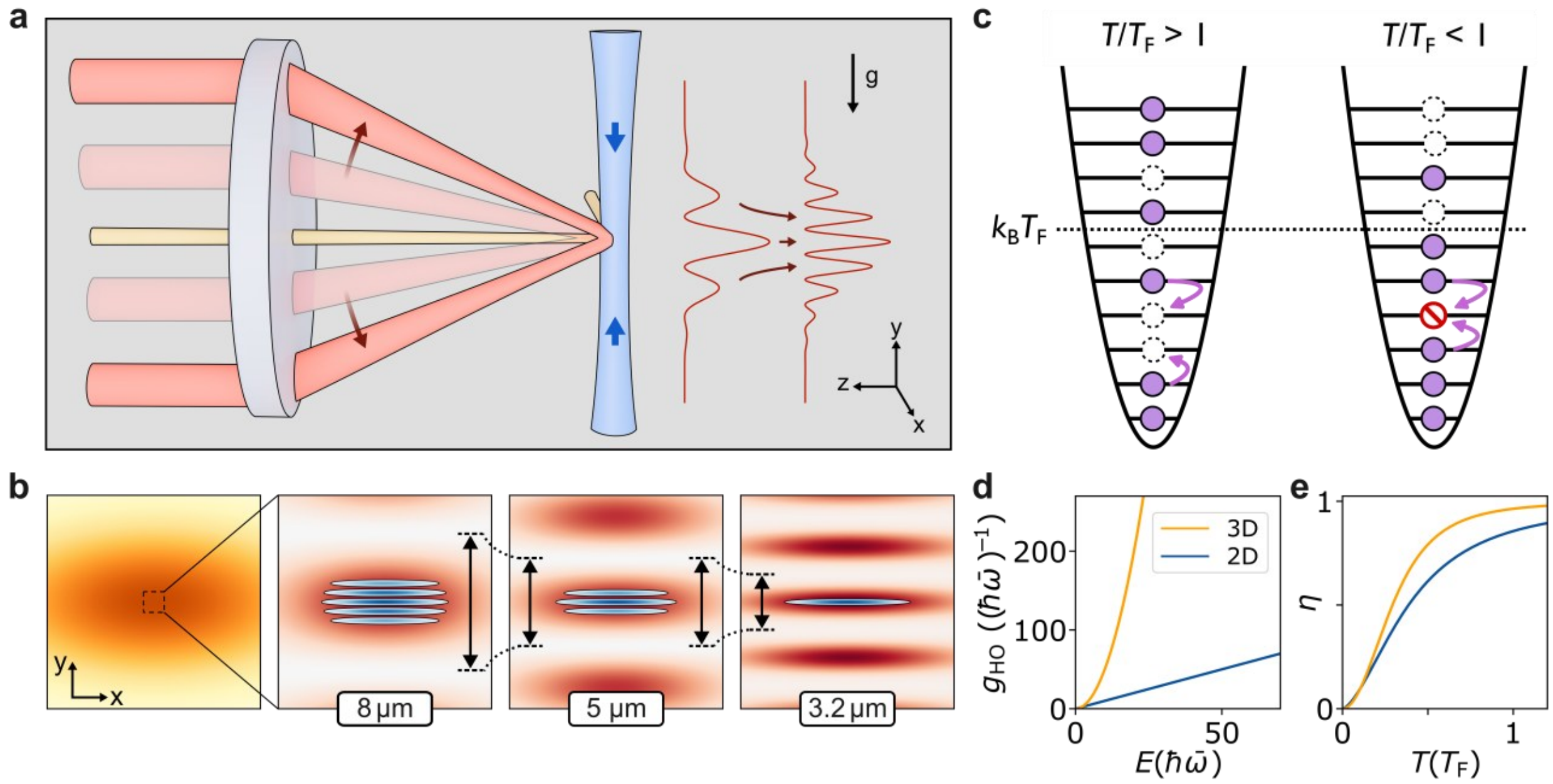

**Fig. 1: Experimental setup and elastic collisions under Pauli blockade.** (**a**) Sketch of the optical trap configurations. The xODT (yellow) and VL (blue) provide the trapping potential for molecules in 3D and quasi-2D, respectively. The AL (red) serves to compress the distribution for the atomic mixture along the y-direction before turning on the VL. The spatial period of the AL intensity distribution (shown as red curves on the right) is adjustable by controlling the intersection angle of AL beams (light red and red) after passing through the focus lens. (**b**) By varying the final spacing of the AL (red) to 8 μm, 5 μm, or 3.2 μm, the atomic mixture – initially trapped in the xODT (yellow) – is loaded into 5, 3, or 1 layer of the VL (blue). (**c**) Elastic collisions between fermionic molecules in harmonic confinement. State-changing collisions are allowed for $T/T_{\mathrm{F}} > 1$ as the final states are mostly unoccupied. Pauli blockade suppresses state-changing collisions for $T/T_{\mathrm{F}} < 1$ as the fraction of unoccupied states are reduced. (**d**) The density of states $g_{\mathrm{HO}}$ for 3D and quasi-2D harmonic trap, where $\bar{\omega}$ is the geometric mean of the trap frequencies. For 3D, $\bar{\omega} = \left(\omega_x \omega_y \omega_z\right)^{1/3}$ while for quasi-2D $\bar{\omega} = (\omega_x \omega_z)^{1/2}$. (**e**) The temperature dependent suppression factor $\eta$ for elastic collisions in 3D and quasi-2D.

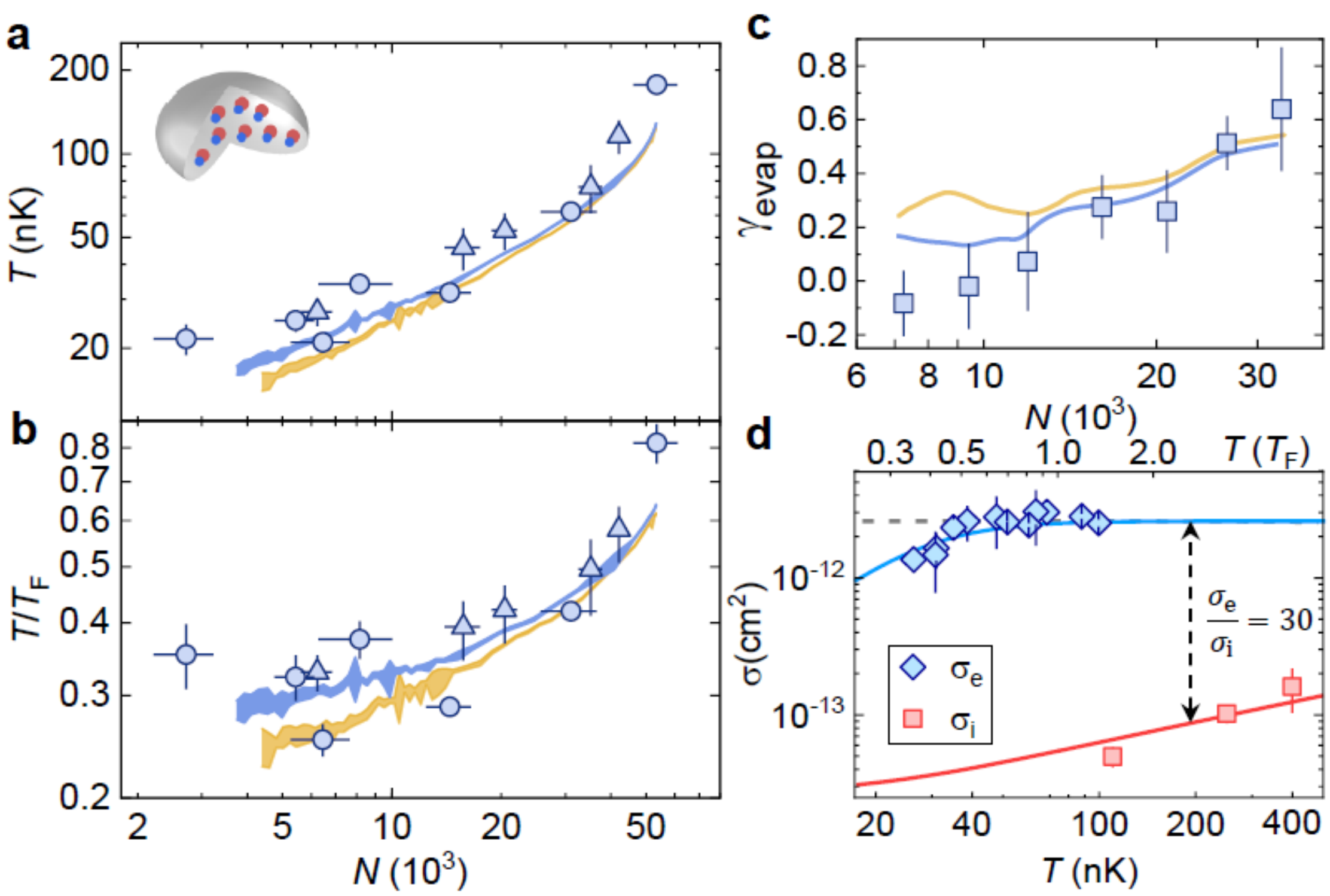

**Fig. 2: Evaporative cooling and collisional cross section measurements of KRb in 3D.** The cooling trajectory for KRb in 3D is shown as $T$ vs $N$ (**a**) and $T/T_F$ vs $N$ (**b**). The circles are from Fermi-Dirac fits while the triangles are from density fluctuation measurements. Error bars are 1 standard error (SE) from Fermi-Dirac fits for the circles and 1 SE of the mean for the triangles (Supplementary Section F). The blue curves are obtained from Monte Carlo simulations of the full evaporation sequence. Yellow curves are obtained from simulations which ignore the Pauli suppression of elastic scattering. The half width of the band represents 1 SE from Fermi-Dirac fits. (**c**) Evaporative cooling efficiency $\gamma_{evap}$ as a function of $N$ in 3D. Values of $\gamma_{evap}$ are extracted from the combined dataset from (a) and (b) using both methods of thermometry. Error bars are 1 SE from fits. Blue and yellow curves correspond to values obtained from the Monte Carlo simulations shown in panels (a) and (b). (**d**) Experimentally measured values of temperature-dependent elastic cross section $\sigma_e$ (blue diamonds) and inelastic cross section $\sigma_i$ (red squares). Error bars are 1 SE from fits. The blue and red curves show the predictions from theory for $T_F =$ 75 nK. The gray dashed line indicates a constant $\sigma_e$ without taking Pauli suppression into account. The ratio $\sigma_e/\sigma_i$ reaches 30 for $T = 200$ nK as the initial temperature of the sample. For an error bar smaller than the size of the marker, the error (1 SE) is represented by half of the gap between two opposing sticks inside the marker.

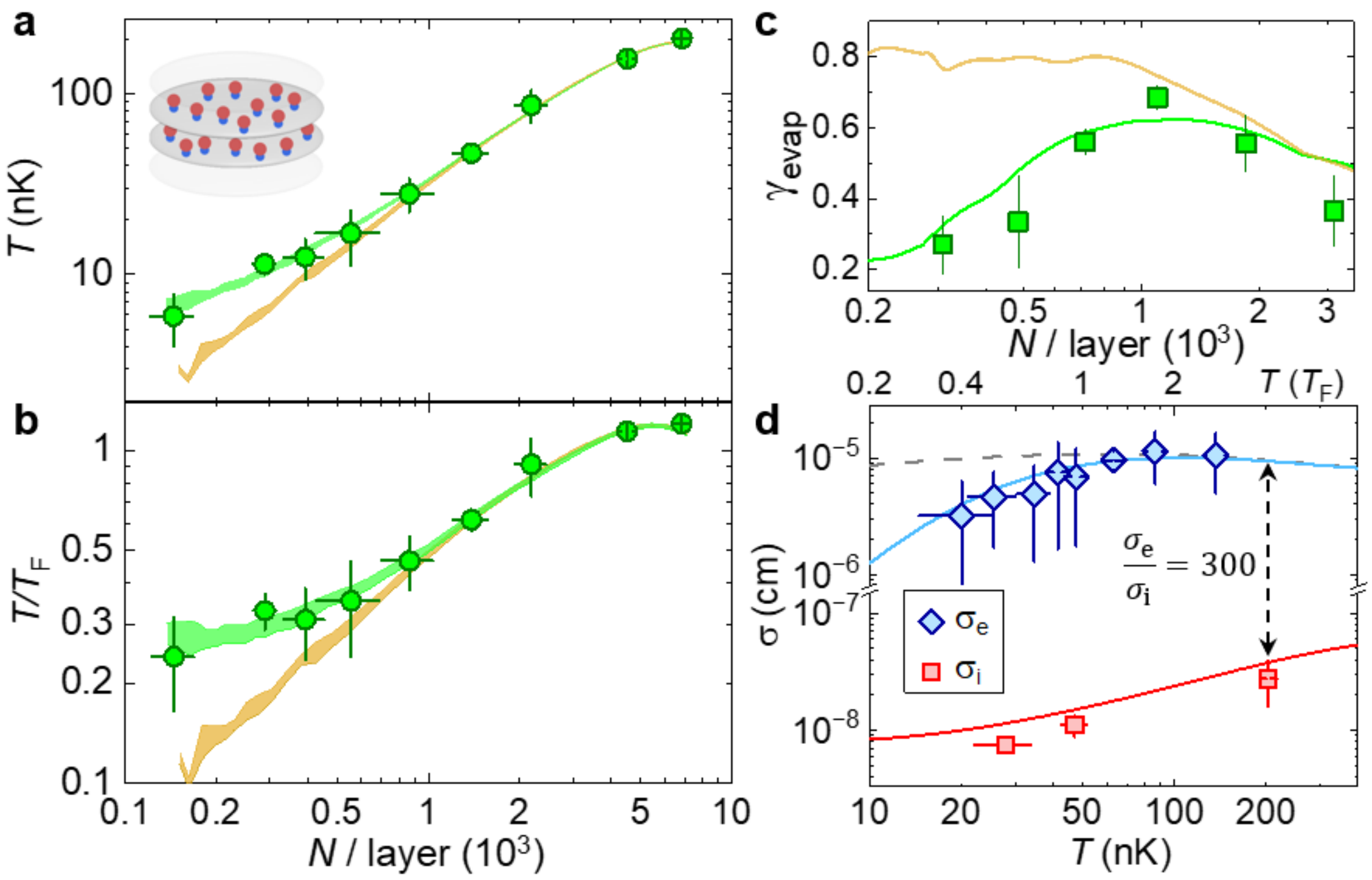

**Fig. 3: Evaporative cooling and collisional cross section measurements of KRb in quasi-2D.** The cooling trajectory for KRb in quasi-2D is shown as $T$ vs $N$ (**a**) and $T/T_F$ vs $N$ (**b**). Data points show our measurement results, obtained from Fermi-Dirac fits to the molecular density profiles. Error bars represent 1 SE of the mean over four experimental repetitions. The green curves are from a Monte Carlo simulation which includes effects from Pauli suppression, laser-induced heating. Yellow curves are from a simulation which excludes Pauli suppression of elastic collisions. The half width of the band represents 1 SE from Fermi-Dirac fits. (**c**) Evaporative cooling efficiency $\gamma_{\text{evap}}$ vs $N$ in quasi-2D. Data points are extracted from the measurements shown in (a), (b). Error bars represent 1 SE from fits. The green and yellow curves are obtained from Monte Carlo simulations as in (a), (b). (**d**) Experimentally measured values of temperature-dependent elastic cross section $\sigma_e$ (blue diamonds) and inelastic cross section $\sigma_i$ (red squares). The blue and red curves show the predictions of the cross sections for $T_F = 50$ nK. The gray dashed line indicates a nearly constant $\sigma_e$ without taking Pauli suppression into account. The ratio $\sigma_e/\sigma_i$ reaches 300 for $T = 200$ nK, the initial temperature of the sample. For an error bar smaller than the size of the marker, the error (1 SE) is represented by half of the gap between two opposing sticks inside the marker.

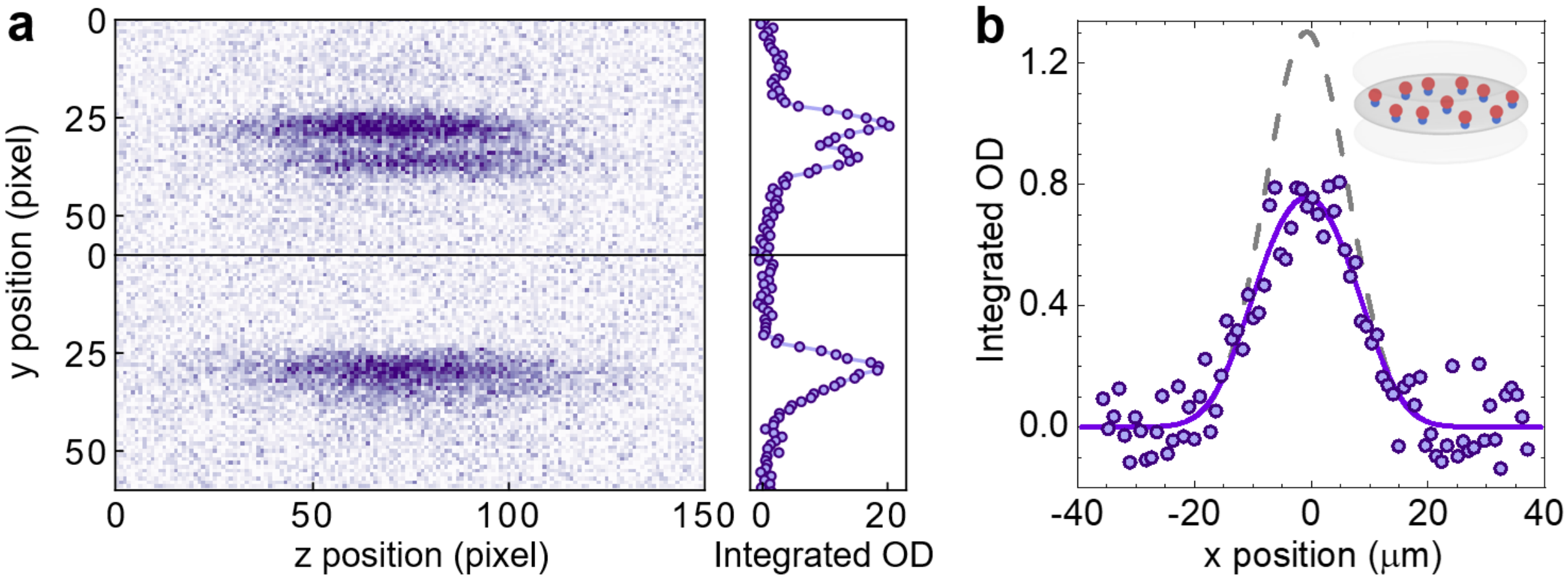

**Fig. 4: Quasi-2D mono- and bilayer samples of ultracold polar KRb.** (**a**) Absorption images (left) and integrated optical density (OD) along the $z$-direction (right) for bilayer (upper panels) and monolayer (lower panels) configurations after matter-wave magnification. (**b**) Integrated OD of a degenerate monolayer KRb cloud (data points). The purple curve shows a fitted Fermi-Dirac distribution, giving $T/T_{\mathrm{F}} = 0.44(14)$. The gray dashed curve is obtained from a Gaussian fit to the outer wings of the sample.

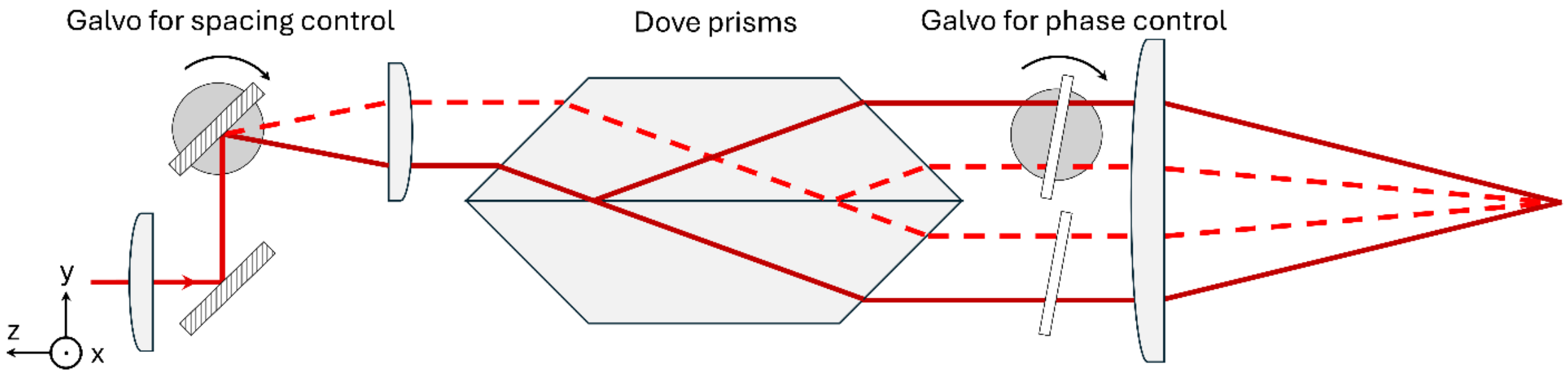

**Fig. 5. The optical layout of the AL.** A single laser beam is split into an upper arm and a lower arm after passing through a pair of dove prisms with a 50:50 beamsplitter coating at their interface. The two arms form a lattice at the focal plane of the last lens where the atoms are held in the xODT. By adjusting the mirror on the first galvo before the dove prism pair, the spacing between the arms can be tuned which subsequently changes the angle of interception at the atoms, realizing a spacing tunable AL. The position of the layers of the AL is adjustable by rotating a phase plate mounted on a galvo which changes the path length of the upper arm but not the lower arm.

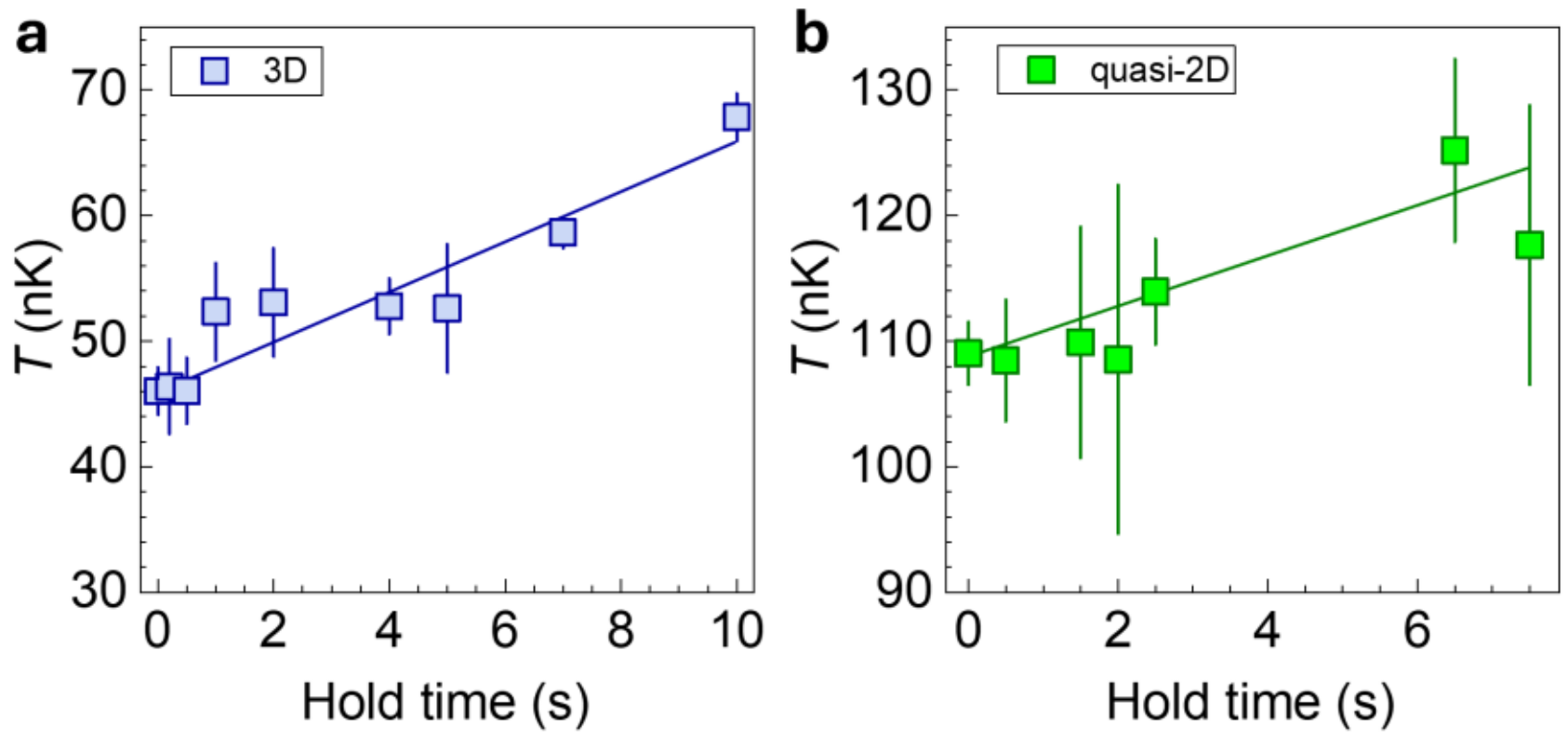

**Fig. 6: One-body heating rate measurements for KRb.** Temperature evolution of KRb (**a**) as a 3D sample holding in the xODT and (**b**) as quasi-2D sample holding in the VL. Error bars are one SE of the mean over five experimental realizations. Solid lines are linear fits to the data. The extracted heating rates from the linear fits are 2.0(2) nK/s and 2.0(4) nK/s for 3D and quasi-2D, respectively.

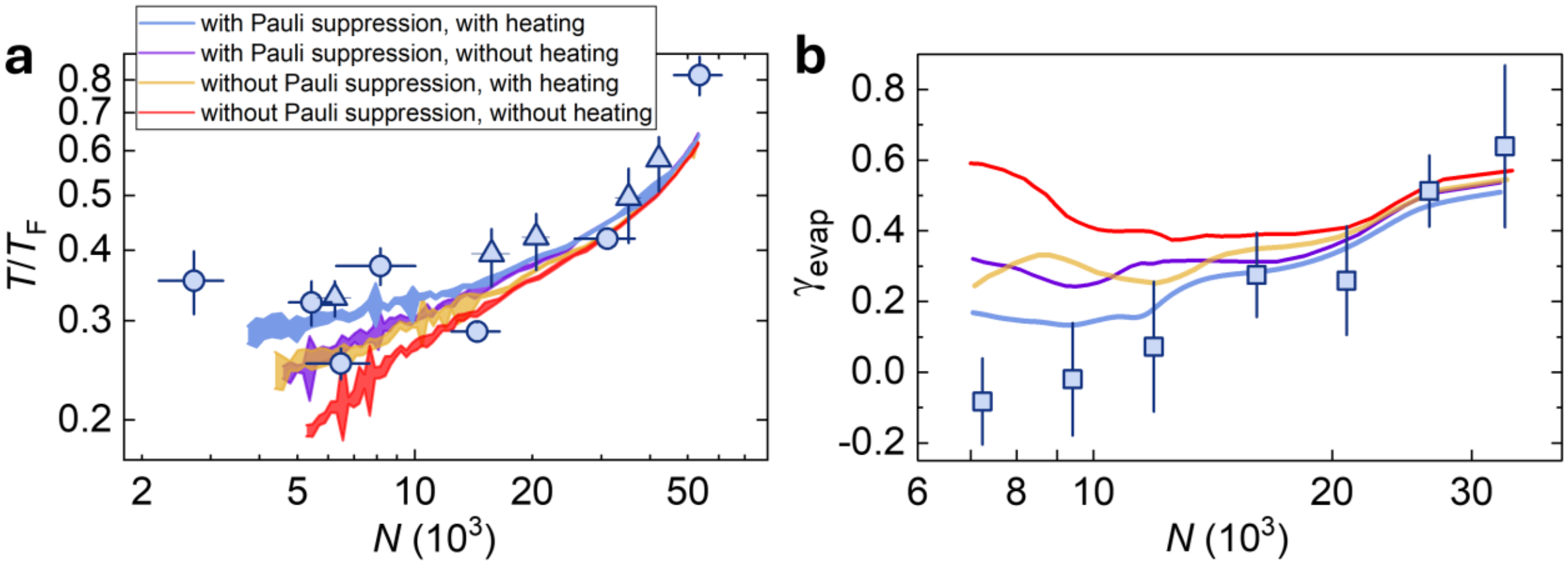

**Fig. 7: Comparison between the measurement and different Monte Carlo simulations in 3D.** (**a**) The measured cooling trajectory (reproduced from Fig. 2b) and Monte Carlo simulations with Pauli suppression and one-body heating (blue), with the Pauli suppression but without one-body heating (purple), without Pauli suppression but with one-body heating (yellow), and without the Pauli suppression and without one-body heating (red). The half width of the band represents 1 SE from Fermi-Dirac fits. (**b**) The cooling efficiency at each stage of the evaporation. The experimental data is reproduced from Fig. 2c. Different colored curves are from Monte Carlo simulations labeled as in (**a**).

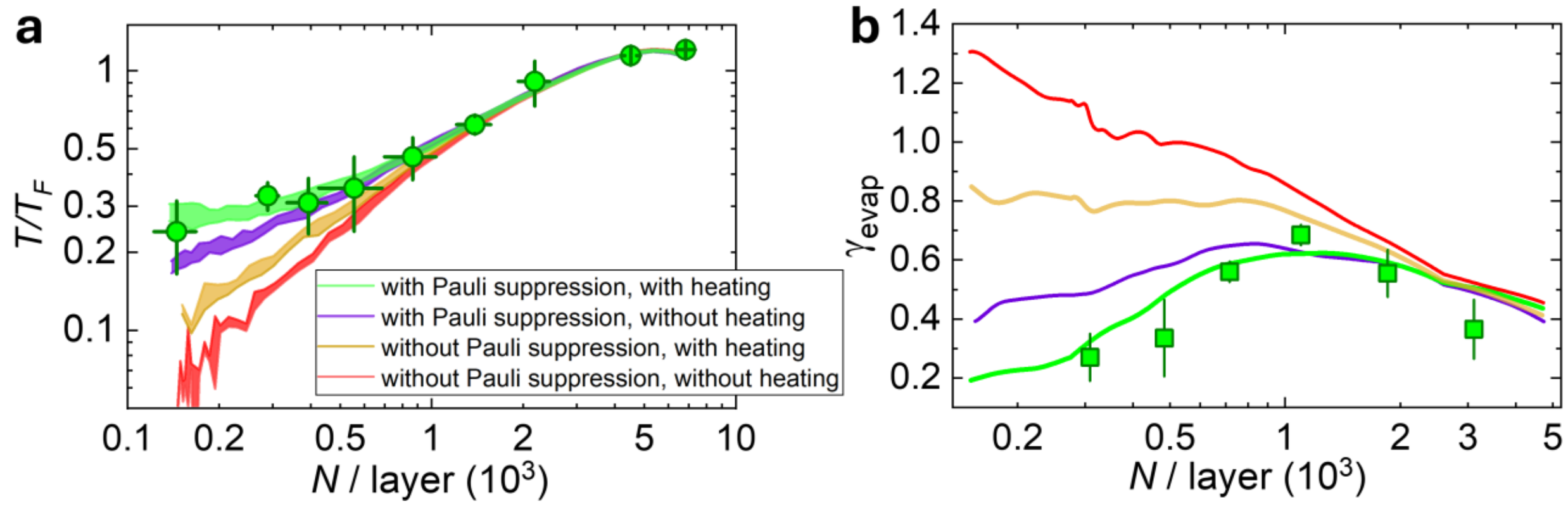

**Fig. 8: Comparison between the measurement and Monte Carlo simulations considering different effects in quasi-2D.** (**a**) The measured cooling trajectory (reproduced from Fig. 3b) and Monte Carlo simulations with Pauli suppression and one-body heating (green), with Pauli suppression but without one-body heating (purple), without Pauli suppression but with one-body heating (yellow), and without Pauli suppression and without one-body heating (red). The half width of the band represents 1 SE from Fermi-Dirac fits. For an error bar smaller than the size of the marker, the error (1 SE) is represented by half of the gap between two opposing sticks inside the marker. (**b**) The cooling efficiency at each stage of the evaporation. The experimental data is reproduced from Fig. 3c. Different colored curves are from Monte Carlo simulations labeled as in (**a**).

# Supplementary Information for

## Fermi gas of polar molecules in the Pauli-blocked regime

Junyu Lin, Annette N. Carroll, Phillip Martin, Calder Miller, Reuben R. W. Wang, Kevin B. Xu, John L. Bohn, Tim de Jongh, Jun Ye

Corresponding authors: Junyu.Lin@colorado.edu, ye@jila.colorado.edu

**Contents**

## A. Matter wave layer distribution imaging

We use matter-wave focusing techniques, which magnify the spatial separation between the different layers, to determine the layer occupation for both the AL and the VL[1–4]. For the AL, we induce a phase-space rotation by instantaneously turning off the lattice potential, followed by a temporal evolution in an auxiliary harmonic potential, provided by the xODT. For an evolution duration of a quarter trap oscillation period, this maps the initial position of the particles onto their momentum. A subsequent TOF expansion allows us to measure this momentum distribution and hence extract the particles' initial position.

While the single phase-space rotation is sufficient to resolve the layer occupation distribution in the AL, which ranges from 3.2 to 8 μm for the sequences used in this work, the small interlayer spacing of 540 nm and tight $\omega_y = 15$ kHz confinement of the VL prevent a reliable measurement of the molecular layer occupation distribution with a single phase-space rotation. Instead, we implement a dual phase-space rotation using two auxiliary harmonic traps, similar to the technique introduced in Ref. 2. This avoids substantial expansion of the intra-layer distribution during TOF that reduces interlayer resolvability for the single phase-space rotation method.

To quantify the system's evolution during the dual phase-space rotation, we consider the initial wave function of the gas along the vertical direction, which is the ground state of a harmonic oscillator with mean position $y$ and momentum $p$ and corresponding variances $\sigma_{y_0}^2 = \hbar/2m\omega_0$ and $\sigma_{p_0}^2 = m\hbar\omega_0/2$. Here, $\omega_0$ is the vertical trap frequency of the initial harmonic potential and $m$ is the particle mass. Upon release into the first auxiliary trap potential with frequency $\omega_1$, the initial phase-space distribution will rotate according to the following rotation matrix:

$$R_{\omega_1}(t) = \begin{bmatrix} \cos\omega_1 t & \frac{1}{\mathrm{m}\omega_1}\sin\omega_1 t \\ -\mathrm{m}\omega_1 \sin\omega_1 t & \cos\omega_1 t \end{bmatrix} \quad (1)$$

Writing position and momentum as a column vector $\vec{u}(t) = [y(t), p(t)]^T$, this matrix can be used to find their values at time $t$ through $\vec{u}(t) = R_{\omega_1}(t)\,\vec{u}(0)$. An evolution of $t = \pi/2\omega_1$ then indeed results in a phase-space rotation of 90°. Rotation of a quarter period in a second auxiliary trap with rotation matrix $R_{\omega_2}(\pi/2\omega_2)$ then results in a spatial magnification of the initial wave function by a factor $M = \omega_1/\omega_2$, since $\omega_1 > \omega_2$. This dual phase-space rotation thus constitutes a matter wave telescope.

We imaged the layer distribution of the VL, using the AL and xODT as the first and second auxiliary traps, respectively. These provided vertical frequencies of $\omega_1 \approx 2\pi \times 2200$ Hz and $\omega_2 \approx 2\pi \times 190$ Hz and hence a spatial magnification of 11.6, resulting in a layer separation of 6.2 μm. With our spatial imaging resolution of 2.1 μm this is marginal for resolving individual layers. We overcome this by slightly deviating from quarter period evolutions in the respective auxiliary traps and adding a TOF evolution. This serves to enlarge the interlayer separation while, as described below, the dual phase-space rotation reduces expansion of the intralayer distribution during TOF, providing enhanced resolvability of the layer occupation distribution.

We can determine the time-evolution of the intralayer position and momentum spreads of the system using the rotation matrix in Eq. 1. The initial spread is described by a diagonal covariance matrix $\mathrm{Cov}[\vec{u}(t)]$ with $\sigma_y^2$ and $\sigma_p^2$ as its elements. After evolution of times $t_1$ and $t_2$ in the first and second auxiliary traps, respectively, the covariance matrix becomes:

$$\mathrm{Cov}[\vec{u}(t_1+t_2)] = R_{\omega_2}(t_2)\, R_{\omega_1}(t_1) \begin{bmatrix} \frac{\hbar}{2m\omega_0} & 0 \\ 0 & \frac{m\hbar\omega_0}{2} \end{bmatrix} R_{\omega_1}^T(t_1)\, R_{\omega_2}^T(t_2) \qquad (2)$$

When the evolution times are exactly set to be the quarter periods of the respective traps, we obtain variances at time $t = t_1 + t_2$ given by $\sigma_y^2(t) = M^2\sigma_{y_0}^2$ and $\sigma_p^2(t) = M^{-2}\sigma_{p_0}^2$, providing a decreased momentum spread.

The spatial spread of the individual layer's vertical wave function during TOF expansion can then be computed from:

$$\sigma_y^2(t_r + t_{\mathrm{TOF}}) = \sigma_y^2(t_r) + \frac{\sigma_p^2(t_r)}{m^2} {t_{\mathrm{TOF}}}^2 + \frac{2t_{\mathrm{TOF}}}{m}\mathrm{Cov}(y,p) \qquad (3)$$

where $t_{\mathrm{r}}$ is the trap-release time and $t_{\mathrm{TOF}}$ is the time of flight. The center-of-mass position for a given layer is simply given by $y(t_{\mathrm{r}} + t_{\mathrm{TOF}}) = y(t_{\mathrm{r}}) + \frac{p(t_{\mathrm{r}})}{m} t_{\mathrm{TOF}}$, with $y(t_r)$ and $p(t_r)$ the position and momentum of the gas at the moment of release. At our typical trap rotation times of $t_1 = 100$ μs $\approx 0.218 \times 2\pi/\omega_1$ and $t_2 = 1.2$ ms $\approx 0.227 \times 2\pi/\omega_2$, and with $t_{\mathrm{TOF}} = 7.2$ ms, we reach interlayer separations of 14 μm while the intralayer spread $\sqrt{\Delta y^2(t_r + t_{\mathrm{TOF}})}$ is maintained below 5 μm. This gives us ample resolution and signal-to-noise ratio (SNR) to detect the layer occupation distribution of the molecules.

We stress that the matter wave interferometry technique described here, consisting of a matter wave telescope intentionally deviating from quarter period evolutions, followed by TOF expansion, can be used to enhance small features of a quantum gas that are otherwise blurred by the limited resolution of the imaging system. The advantage compared to a single phase-space rotation and to a matter wave telescope with quarter period evolutions lies in the ability to enhance spatial features beyond the magnification set by the auxiliary harmonic traps while maintaining sufficient SNR. As described above, this advantage arises from the TOF expansion in concurrence with the demagnification the dual phase-space rotation provides for the momentum spread of the gas.

### B. Layer distributions for bilayer and monolayer configurations

By applying the matter wave focusing technique introduced in the previous section, we measured the populations of KRb molecules in the VL for both bilayer and monolayer configurations.

For the bilayer configuration, the relative populations of the two layers were 0.58 and 0.42 with a standard deviation of 0.09, as shown in Supplementary Fig. 1a, obtained by averaging over forty experimental repetitions. Over all these realizations, the relative population in each layer was quite stable, as illustrated by the small spread in the histogram shown in Supplementary Fig. 1b. The routinely near-equal population distribution demonstrates our ability to reliably and reproducibly prepare molecules in two layers of the VL.

Such a small population imbalance in the two layers is not expected to significantly affect the thermometry. Because the absorption imaging is performed along the $y$-direction, the OD is integrated over the gas in both layers. As a result, the reduced temperature obtained from Fermi-Dirac fits represents a population-weighted average of the reduced temperatures in the two layers. Assuming the molecules in both layers are at the same temperature, the extracted reduced temperature should deviate no more than 10% from that of either layer individually.

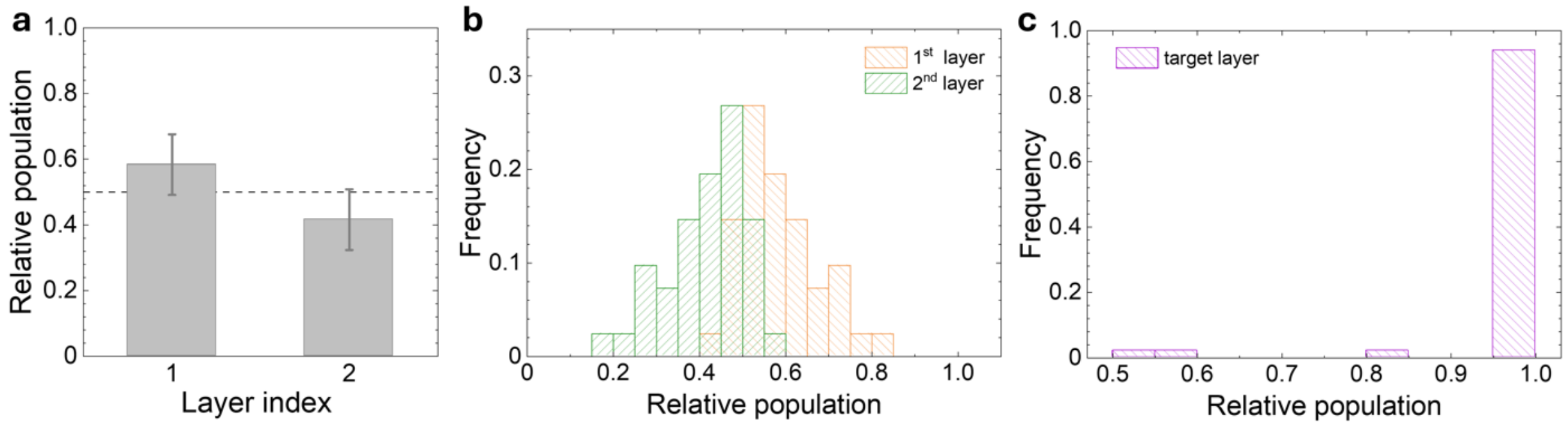

**Supplementary Fig. 1: Distributions of molecules in the VL.** (**a**) Relative population of molecules between layers in the VL after the bilayer preparation sequence. Error bars represent one standard deviation of the mean over forty experimental realizations. The dashed line shows 0.5 relative population to guide the eye. (**b**) Distribution of the relative populations in the 1st layer and the 2nd layer for the bilayer configuration. (**c**) Distribution of the relative populations in the target layer for the monolayer configuration.

For the monolayer configuration, in addition to the strong evidence of successful single-layer loading provided by the matter-wave-refocused images of KRb molecules shown in Fig. 4a of the main text, we obtained further statistics suggesting routinely successful preparation of monolayers by imaging the VL layer distribution of Rb over many experimental realizations. For the quantitative analysis of the monolayer configuration, matter-wave focusing was applied to the atomic mixture immediately prior to molecule association, instead of directly imaging the molecules. This method enhanced the detection of small populations in any undesired layers, as the Rb BEC retains a relatively high OD after TOF expansion compared to the KRb molecules.

Across a total of fifty measurements, we observed forty-seven instances in which the population occupied a single VL layer, while only three instances exhibited a noticeable population in an unwanted second layer, as shown in Supplementary Fig. 1c. This corresponds to a 94% success rate for single-layer preparation. If desired for future measurements, the residual population in the unwanted layer could be removed by performing layer selection directly on the KRb molecules as previously demonstrated in Ref. 5.

**C. Evaporation cooling trajectory of monolayer**

The evaporative cooling trajectory of molecules in the monolayer configuration is presented in Supplementary Fig. 2, showing that the molecules experience a similar evaporation efficiency of an overall $\gamma_{\mathrm{evap}} = 0.34(2)$ in the monolayer configuration compared to the bilayer configuration. As presented in the main text, after evaporating to around 200 molecules, we achieve a $T/T_{\mathrm{F}} = 0.44(14)$. Since the molecules have a similar cooling efficiency in both the monolayer and bilayer configurations, we attribute the higher final $T/T_{\mathrm{F}}$ of the monolayer to slightly worse initial conditions compared to the bilayer configuration. The monolayer has a higher initial $T/T_{\mathrm{F}}$ of 1.40(1) compared to 1.21(1) in the bilayer configuration because of additional heating from the extra power compression of the AL that is needed to load the atoms into a single layer of the VL. Additional efforts to reduce nonadiabatic heating during the additional compression should enable preparation of a lower initial $T/T_{\mathrm{F}}$, which should result in a $T/T_{\mathrm{F}} \approx 0.25$ after evaporation as achieved in the bilayer configuration.

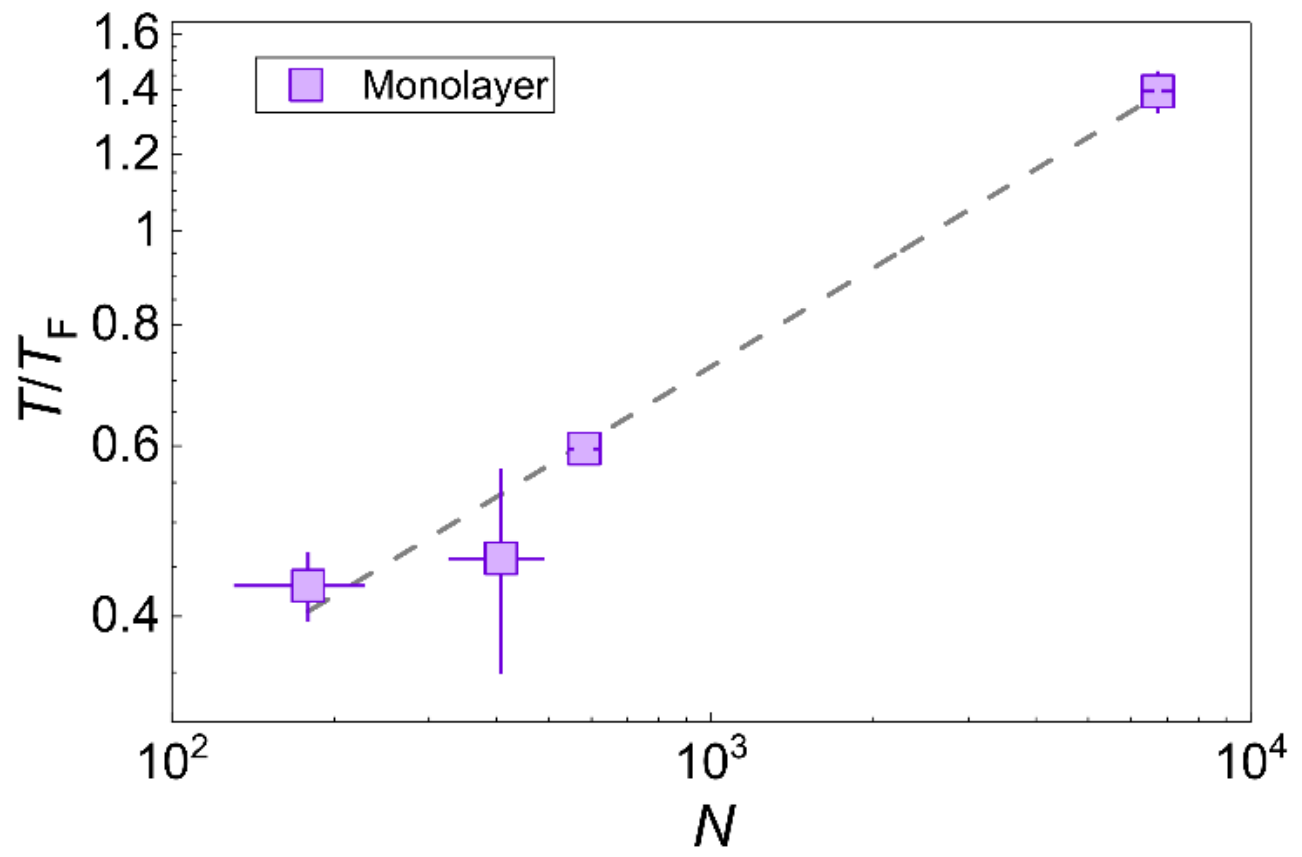

**Supplementary Fig. 2: Evaporative cooling trajectory for monolayer KRb.** Error bars represent one standard error (SE) of the mean over three experiment realizations. The dashed line is a linear fit with its slope corresponding to an overall $\gamma_{\text{evap}} = 0.34(2)$. For an error bar smaller than the size of the marker, the error (1 SE) is represented by half of the gap between two opposing sticks inside the marker.

**D. Interference fringe removal**

At low ODs, the raw experimental images of the quasi-2D gas taken along the vertical imaging direction display interference fringes which are even more pronounced when the image is averaged over many experimental realizations. To mitigate the effect of background patterns on the fitted molecular density profiles, we apply a well-established background-removal algorithm [6–8].

In brief, for each absorption image we construct a column vector consisting of pixel values in the region of the image not containing molecules. This background region is typically comprised of over 4000 pixels. We accordingly compose such vectors for a set of images not containing any molecules in the entire image, which we refer to as our reference image basis set. To remove fringes from a given image, we use a standard least-square algorithm to optimally project the background region of this image onto our basis set. This way we reconstruct the background region of the absorption image as a weighted sum of reference images from our basis set. Next, the weighted sum of reference images is subtracted from the full absorption image, such that fringes inside the region with particles are also removed. Results of the background removal, shown in Supplementary Fig. 3a, show a clear reduction of the interference fringes.

We determine the minimally required reference basis set size by applying the background removal algorithm to a fixed set of experimental images for a range of reference basis set sizes and evaluating the convergence of the signal-noise ratio SNR of the resulting image. For a given basis set size, we obtain the SNR by fitting the molecular density profile, setting the fitted peak amplitude as the signal and the standard deviation of the fit residual as the noise. Resulting SNRs are plotted against basis set size in Supplementary Fig. 3b for an image averaged over 20 experimental realizations with a peak OD of about 0.35. For images taken at the later steps in our evaporation sequence, where we have peak ODs of $\lesssim 0.1$, a reference basis set size of about 100 images is required.

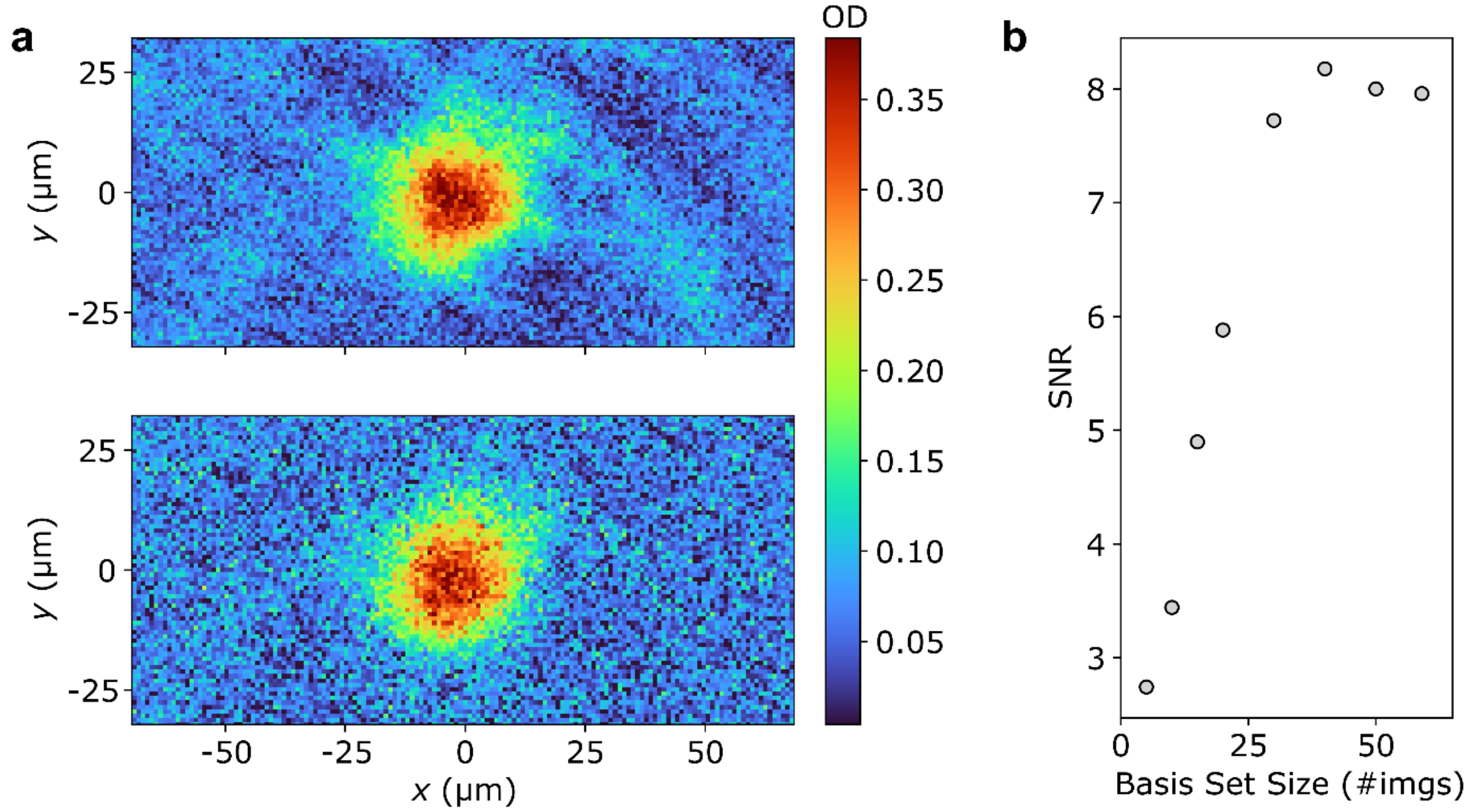

**Supplementary Fig. 3: Interference fringe removal in the OD profiles of quasi-2D Fermi gases obtained through vertical imaging. (a)** Raw OD profile averaged over 20 experimental realizations (top) and OD profile of the same data after applying the fringe removal algorithm (bottom). **(b)** SNR of the OD profiles after fringe removal as a function of basis set size used for the background removal. Fitted widths and reduced temperatures display a similar convergence with basis set size.

### E. Thermometry by image fitting

For all experiments, the molecule number and temperature were extracted by fitting a Fermi-Dirac density distribution[9] to the absorption images of the molecular cloud.

For a Fermi gas in a 3D harmonic trap, the density distribution integrated along the $x$-direction is

$$n_{3D}(y,z) = -n_0 \mathrm{Li}_2\left(-\zeta e^{-\frac{1}{2}\left(\frac{y^2}{\sigma_y^2}+\frac{z^2}{\sigma_z^2}\right)}\right). \tag{4}$$

Here, $\mathrm{Li}_n$ is the $n^{\mathrm{th}}$ order polylogarithm, and $n_0$, $\zeta$, $\sigma_y$ and $\sigma_z$ are discussed below and are treated as fitting parameters.

The reduced temperature $T/T_\mathrm{F}$ of the cloud in a 3D harmonic trap is directly linked to the fitted fugacity $\zeta$ through

$$\left(\frac{T}{T_\mathrm{F}}\right)^3 = -\frac{1}{6\mathrm{Li}_3(-\zeta)}\ . \tag{5}$$

The temperature $T_i$ along the $i$-direction can also be extracted from the fitted size $\sigma_i$, according to

$$\sigma_i = \frac{\sqrt{1+\omega_i^2 t_\mathrm{TOF}^2}}{\omega_i}\sqrt{\frac{k_\mathrm{B} T_i}{m}}, \tag{6}$$

with $\omega_i$ the trap frequency along the $i$-direction ($i = y, z$) and $t_\mathrm{TOF}$ the time of flight.

For a quasi-2D Fermi gas in a harmonic trap, the density distribution integrated along the y-direction is

$$n_{2D}(x,z) = -n_0 \mathrm{Li}_1\left(-\zeta e^{-\frac{1}{2}\left(\frac{x^2}{\sigma_x^2}+\frac{z^2}{\sigma_z^2}\right)}\right). \tag{7}$$

To enhance the SNR for the gas in quasi-2D, we typically additionally integrated the density along the $z$-direction and fitted with an integrated density distribution

$$n_{2D,\mathrm{int}}(x) = -n_0 \mathrm{Li}_{\frac{3}{2}}\left(-\zeta e^{-\frac{x^2}{2\sigma_x^2}}\right) \tag{8}$$

The reduced temperature $T/T_\mathrm{F}$ of the cloud in a 2D harmonic trap is obtained from the fitted fugacity according to

$$\left(\frac{T}{T_\mathrm{F}}\right)^2 = -\frac{1}{2\mathrm{Li}_2(-\zeta)}, \tag{9}$$

and $T_x$ can also be extracted from Eq. 6. The reduced temperature $T_x/T_\mathrm{F}$ can additionally be obtained from the extracted $T_x$ and the calculated quasi-2D Fermi temperature, $T_\mathrm{F} = \hbar\bar{\omega}\sqrt{2N}/k_\mathrm{B}$, with the experimentally measured $\bar{\omega}$ and the fitted $N$.

Supplementary Fig. 4 shows a comparison between the two methods for extracting the reduced temperature: one by extracting $T/T_\mathrm{F}$ from the fitted fugacity and the other by extracting $T_x/T_\mathrm{F}$ from the fitted $\sigma_x$ and calculated $T_\mathrm{F}$. We find very good agreement between the two methods of extracting reduced temperature in quasi-2D.

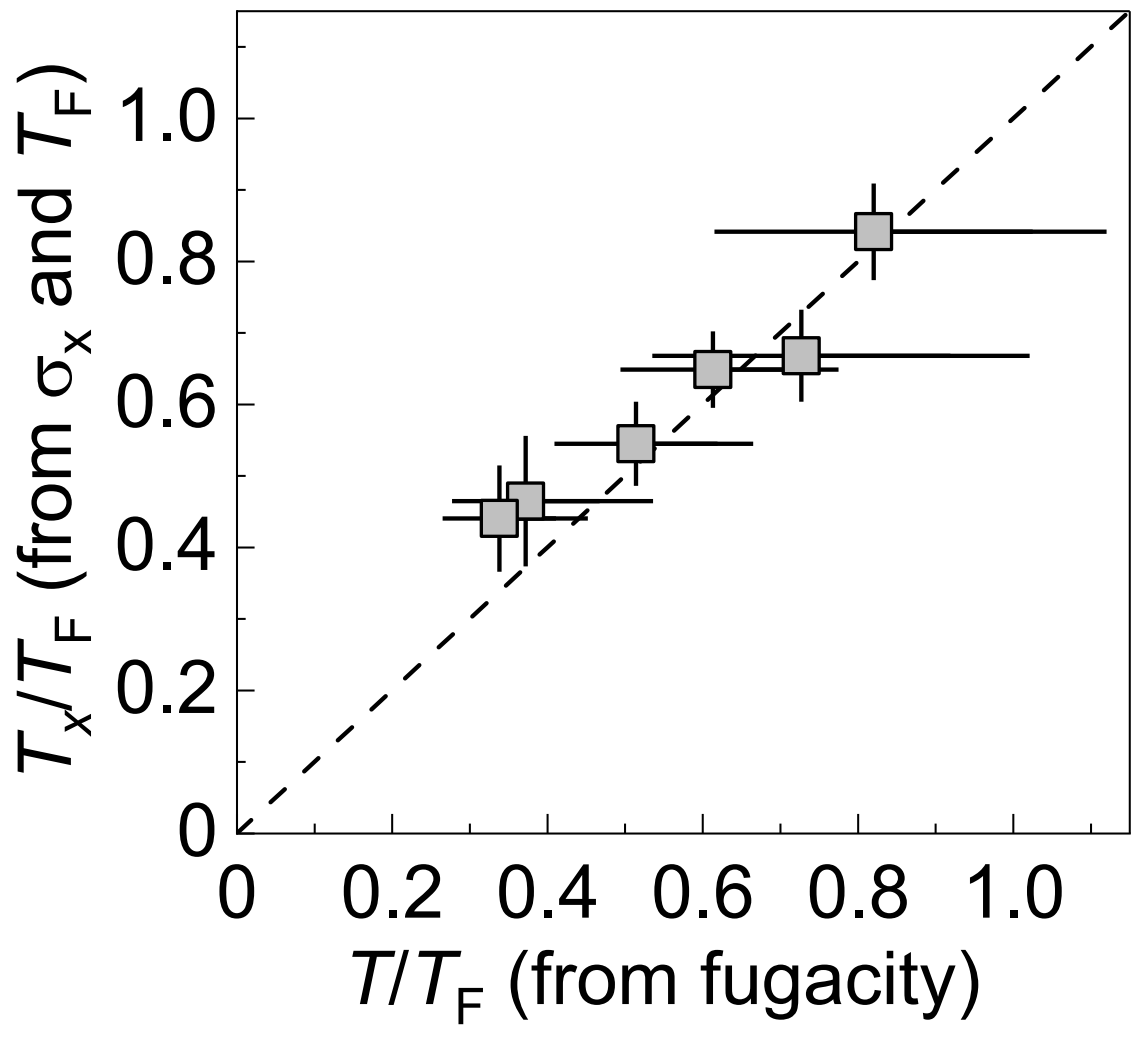

**Supplementary Fig. 4: Thermometry comparison for KRb in quasi-2D.** $T_x/T_\mathrm{F}$ extracted from the fitted $\sigma_x$, $t_\mathrm{TOF}$ and $T_\mathrm{F}$ is plotted against $T/T_\mathrm{F}$ extracted from the fitted fugacity. The error bars are 1 SE from Fermi-Dirac fits. The dashed line indicates equal $T_x/T_\mathrm{F}$ and $T/T_\mathrm{F}$.

For molecules in 3D, due to anharmonic defects in the trap potential, we find that our fugacity-derived $T/T_\mathrm{F}$ is 40% of the value of the size-derived $T_z/T_\mathrm{F}$ when we fit our images with a Fermi-Dirac profile derived for a perfect harmonic trap. We developed a fitting model that considers an anharmonic trapping potential in the central 6-μm-radius region, improving the agreement between the two reduced temperatures to $T/T_\mathrm{F} \approx 0.7\ T_z/T_\mathrm{F}$. The reported temperatures in Fig. 2 are extracted from this improved anharmonic model. To have an independent measurement of $T/T_\mathrm{F}$ in 3D, we additionally perform density fluctuations thermometry, as introduced in the next section.

## F. Density fluctuation thermometry

Measurements of density fluctuations provide an additional thermometry method that is less dependent on the trap geometry along the imaging plane. Here, the mean local molecule number $\bar{N}$ within a subregion of the gas is compared to the variance within this region, Var $N$, obtained over many experimental realizations. While a thermal gas will display Poissonian fluctuations, for which Var $N = \bar{N}$, a degenerate Fermi gas will display a reduced number variance arising from Fermi suppression of density fluctuations. These fluctuations are extractable from both *in-situ* and time-of-flight absorption images, as demonstrated experimentally[10–14].

We perform density fluctuation thermometry of the 3D gas along the tightly confined y-direction (vertical) for which we take *in-situ* images, as well as along the x-direction (side), for which we take absorption images after 4.4 ms of TOF. To extract density fluctuations, we follow the procedure described in Ref. 13. In brief, we post-select images by allowing a 10-20% deviation from the mean global molecule number and subdivide our images into square pixel bins. We subsequently fit each image with a Fermi-Dirac distribution and subtract it from the experimental image to remove effects from global number fluctuations on the local density variance. Next, we take the mean and variance of the molecule number in each bin of the resulting images over 40 to 70 individual experimental realizations and subtract contributions from photon shot noise. We additionally correct for fluctuations introduced by our finite STIRAP efficiency at 12.7 kV/cm of 85%. Finally, to account for variance reductions arising from the imaging system's finite resolution and depth-of-field[13,15], we scale the resulting variance by a factor determined by scaling fluctuations obtained for a thermal gas to the Poissonian case (Var $N = \bar{N}$).

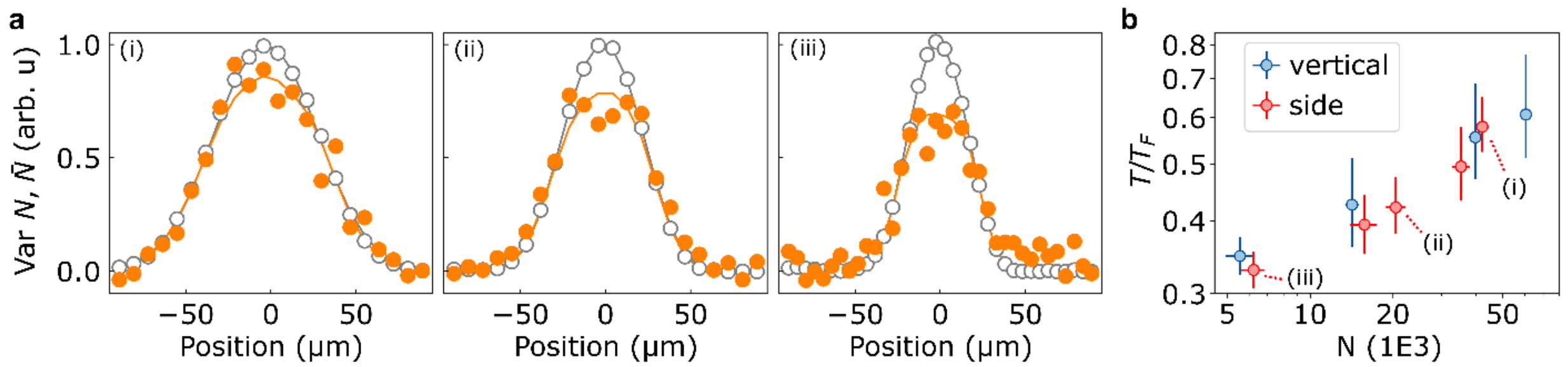

**Supplementary Fig. 5: Density fluctuations thermometry of a three-dimensional gas of degenerate KRb molecules throughout the evaporation trajectory. (a)** Mean number profiles (open grey circle) are shown together with the extracted variance profiles (orange circles) at several points throughout the evaporation trajectory, as indicated by the labels (i)-(iii). The ratio between the mean and variance profiles are used to extract the reduced temperatures shown in panel (b). Fitted mean and variance profiles are given by the grey and orange lines, respectively. **(b)** $T/T_F$ extracted from fluctuation thermometry plotted against the mean global number throughout of the evaporation trajectory. Error bars represent the standard error of the mean obtained from a set of $Var\ N/N$ values in the central region of the cloud. The data points corresponding to the profiles shown in (a) are indicated by the respective labels.

Resulting mean and variance profiles are shown in Supplementary Fig. 5a at several steps throughout the evaporation trajectory. As the molecules evaporate to lower temperatures, the ratio Var $N/\bar{N}$ is seen to manifestly decrease, as is especially apparent in the central region of the gas, where the local chemical potential is maximal. This suppression of density fluctuations constitutes a clear sign of Fermi degeneracy.

We obtain the reduced temperatures $T/T_{\mathrm{F}}$ from these measurements by relating the local density fluctuations of the gas to its global chemical potential and temperature. In the thermodynamic limit and within the Local Density Approximation, the local variance of the density $\mathrm{Var}\, n(\vec{r})$ at position $\vec{r}$ can be related to the mean local density $\bar{n}(\vec{r})$ and temperature $T$ through the fluctuation-dissipation theorem[13]. For a non-interacting fermionic system this yields:

$$\frac{\mathrm{Var}\, n(\vec{r})}{\bar{n}(\vec{r})} = \frac{\mathrm{Li}_{\frac{1}{2}}\left(-e^{\beta(\mu_0 - V(\vec{r}))}\right)}{\mathrm{Li}_{\frac{3}{2}}\left(-e^{\beta(\mu_0 - V(\vec{r}))}\right)} \tag{10}$$

Where $\mu_0$ is the global chemical potential, $\beta = 1/k_B T$, $V(\vec{r})$ the trapping potential. We relate this to the experimentally accessible column densities, by integrating Eq. 10 along the imaging direction, which we here set to be $x$. Assuming the potential is harmonic along the $x$-axis, we find an expression for the local number fluctuations within a given spatial bin on the absorption image in which the potential is assumed to be approximately constant:

$$\frac{\mathrm{Var}\, N}{\bar{N}} = \frac{\mathrm{Li}_1\left(-e^{\beta(\mu_0 - V(y,z))}\right)}{\mathrm{Li}_2\left(-e^{\beta(\mu_0 - V(y,z))}\right)} \tag{11}$$

with $V(y, z)$ the potential along the directions of the image plane.

To avoid systematic errors caused by the deviation from a purely harmonic trap, we take the average ratio $\frac{\mathrm{Var} N}{\bar{N}} = \frac{\mathrm{Li}_1(-e^{\beta\mu_0})}{\mathrm{Li}_2(-e^{\beta\mu_0})}$ in the central region of the cloud – where $V(y, z) \approx 0$ – and obtain $\mu_0$. Any variation of the trap potential in the imaging plane can then only lead to an overestimation of this ratio and thus to an underestimation of $\beta\mu_0$. We subsequently determine the global reduced temperature $T/T_{\mathrm{F}}$ using:

$$\frac{T}{T_{\mathrm{F}}} = \left[-6\mathrm{Li}_3\left(-e^{\beta\mu_0}\right)\right]^{-\frac{1}{3}} \tag{12}$$

Resulting values throughout the evaporation trajectory are shown in Supplementary Fig. 5b, showing excellent agreement between results from the two imaging axes. We also find excellent agreement with the reduced temperatures obtained from the density profile fits, and we show the values obtained with both methods in Fig. 2b of the main text.

Since Eq. 12 strictly holds for a harmonic potential, we numerically determine the expected systematic error on the extracted reduced temperatures arising from anharmonicities in the central region of our trap. Deviations from harmonicity in the imaging plane are numerically determined to underestimate $T/T_{\mathrm{F}}$ by 7-12.5%, whereas deviations from harmonicity in the imaging direction lead to an underestimation of at most 10%. The excellent agreement obtained for the $T/T_{\mathrm{F}}$ values from fluctuation thermometry along both imaging axes therefore indicates these systematic uncertainties are likely on the order of our statistical uncertainties.

### G. Measuring the elastic cross sections

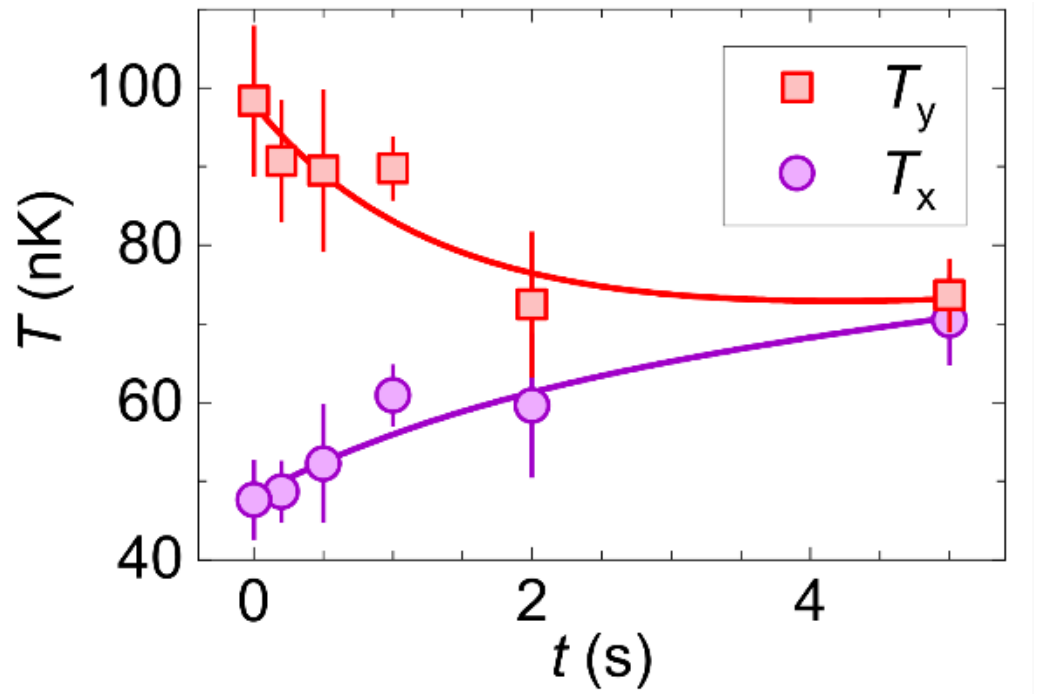

**Supplementary Fig. 6: Example of CDT measurement for KRb in 3D.** Temperature evolution after trap perturbation which initially increases the temperature in the y-direction. Error bars represent one SE of the mean over five experimental realizations. Solid curves are fits from Eq. 13-16.

To measure the elastic cross sections in 3D, we performed cross-dimensional thermalization (CDT) measurements by inducing a temperature imbalance between the $y$-direction and the $x$- and $z$-directions. An example CDT measurement is shown in Supplementary Fig. 6. The temperature evolution is fitted to a set of equations [16,17]:

$$\frac{\mathrm{d}T_y}{\mathrm{d}t} = \frac{n}{12}K_{\mathrm{L,3D}}(T_x + T_z - T_y)T_y - \eta\left(\frac{n\sigma_{\mathrm{e,th}}v_r}{3\mathrm{N}_{\mathrm{col,yx}}}(T_y - T_x) + \frac{n\sigma_{\mathrm{e,th}}v_r}{3\mathrm{N}_{\mathrm{col,yz}}}(T_y - T_z) + \frac{\gamma_{\mathrm{e}}}{3}(T_y - T_x) + \frac{\gamma_{\mathrm{e}}}{3}(T_y - T_z)\right) + c_y \quad (13)$$

$$\frac{\mathrm{d}T_x}{\mathrm{d}t} = \frac{n}{12}K_{\mathrm{L,3D}}(T_y + T_z - T_x)T_x - \eta\left(\frac{n\sigma_{\mathrm{e,th}}v_r}{3\mathrm{N}_{\mathrm{col,yx}}}(T_x - T_y) + \frac{n\sigma_{\mathrm{e,th}}v_r}{3\mathrm{N}_{\mathrm{col,xz}}}(T_x - T_z) + \frac{\gamma_{\mathrm{e}}}{3}(T_x - T_y) + \frac{\gamma_{\mathrm{e}}}{3}(T_x - T_z)\right) + c_x \quad (14)$$

$$\frac{\mathrm{d}T_z}{\mathrm{d}t} = \frac{n}{12}K_{\mathrm{L,3D}}(T_x + T_y - T_z)T_z - \eta\left(\frac{n\sigma_{\mathrm{e,th}}v_r}{3\mathrm{N}_{\mathrm{col,yx}}}(T_z - T_x) + \frac{n\sigma_{\mathrm{e,th}}v_r}{3\mathrm{N}_{\mathrm{col,zy}}}(T_z - T_y) + \frac{\gamma_{\mathrm{e}}}{3}(T_z - T_y) + \frac{\gamma_{\mathrm{e}}}{3}(T_z - T_x)\right) + c_z \quad (15)$$

$$\frac{\mathrm{d}N}{\mathrm{d}t} = -\frac{1}{3}K_{\mathrm{L,3D}}(T_x + T_y + T_z)nN - \gamma_{\mathrm{one-body,3D}}N. \quad (16)$$

Here, the fitting parameter $\eta$ is the Pauli-suppression factor for elastic scattering, $K_{\mathrm{L}}$ describes the two-body loss, and $c_x$, $c_y$ and $c_z$ are one-body heating rates for the $x$-, $y$-, and $z$-directions, respectively. To reduce the number of fitting parameters, we set $c_x = c_z$. Furthermore, $\sigma_{\mathrm{e,th}}$ is fixed to $2.58 \times 10^{-12}\mathrm{cm}^2$ as the elastic cross section for a thermal ensemble which is calculated from theory and agrees well with the measured value at high temperatures. $\gamma_e = 0.21(5)\ \mathrm{s}^{-1}$ is the thermalization rate caused by the anharmonicity of the trap and $\gamma_{\mathrm{one-body,3D}} = 0.056(4)\ \mathrm{s}^{-1}$ is the one-body loss rate of KRb molecules due to off-resonant scattering with photons from the trapping laser, which are both measured independently. The number of collisions required for thermalization are calculated from theory[18], with $\mathrm{N_{col,xy}} = 4.0$, $\mathrm{N_{col,yz}} = 1.3$, and $\mathrm{N_{col,xz}} = 9.0$ for thermalization between the $x$-$y$, $y$-$z$, and $x$-$z$ directions. Finally, $n$ and $v_r$ are the mean density and mean relative collision velocity, respectively, of the sample with Fermi-Dirac density distributions.

This model captures four main contributions for the CDT: (1) the thermalization due to the anisotropic elastic dipolar collisions; (2) the thermalization due to the anharmonicity of the trap potential; (3) the suppression of (1) and (2) from Pauli blockade; (4) the heating from two-body loss and technical noise. The effect from the trapping anharmonicity is expected to be suppressed at lower temperatures[19]. We assume such a factor approximately equals to $\eta$ over the range of measured temperatures, the equations are further simplified to the current forms of Eq. 13-16 used for fitting. The ensemble-averaged elastic cross-section is reported as $\sigma_\mathrm{e} = \eta\sigma_\mathrm{e,th}$ in the main text.

For KRb in quasi-2D, due to the large dipole moment, the temperature evolution is dominated by elastic collisions. Thus, the equations for fits are further simplified to

$$\frac{\mathrm{d}T_x}{\mathrm{d}t} = -\frac{n\sigma_\mathrm{e}v_r}{2\mathrm{N_{col,2D}}}(T_x - T_z) \tag{17}$$

$$\frac{\mathrm{d}T_z}{\mathrm{d}t} = -\frac{n\sigma_\mathrm{e}v_r}{2\mathrm{N_{col,2D}}}(T_z - T_x). \tag{18}$$

The number of collisions required for thermalization in quasi-2D, $\mathrm{N_{col,2D}}$, is calculated from theory with a value between 5 to 7 depending on the temperature[20]. The number of KRb is nearly constant throughout the CDT measurement.

**H. Measuring the inelastic cross sections**

The inelastic cross sections in 3D are measured by fitting the number evolutions of KRb in the xODT to Eq. 16. We prepared a high-density sample to perform this measurement so that the loss is dominated by the two-body loss instead of the one-body loss due to the off-resonant scattering of the trapping laser. With the fitted $K_\mathrm{L,3D}$, the inelastic cross section can be extracted as $\sigma_\mathrm{i} = K_\mathrm{L,3D}T/v_r$ with $T$ the mean temperature and $v_r$ the mean relative collision velocity.

The measurements for inelastic cross sections in quasi-2D is similar to 3D, where we extract $\sigma_\mathrm{i} = K_\mathrm{L,2D}T/v_r$ from fitting the number evolution in VL to a two-body loss equation $\frac{\mathrm{d}N}{\mathrm{d}t} = -K_\mathrm{L,2D}TnN - \gamma_\mathrm{one-body,2D}N$. The one-body loss rate in VL $\gamma_\mathrm{one-body,2D}$ is expected to be 0.016 $\mathrm{s^{-1}}$ (Ref. 21) which is negligible.

## I. Evaporative cooling sequence

For evaporative cooling of KRb molecules in 3D, we decreased the xODT laser power in linear ramps that interpolated an exponential shape. The trap depth $U$ of the xODT for the optimized sequence as a function of evaporation time is shown as Supplementary Fig. 7a. Together with the measured temperature of the gas, we show the ratio $\frac{U}{k_B T}$ at each step of the evaporation in Supplementary Fig. 7b. The $\frac{U}{k_B T}$, initially around 12 for molecules stably trapped in the xODT, is reduced to around 5 throughout the evaporation sequence, which is smaller than the typical value of 8 for evaporative cooling of atoms[22]. We attribute the optimal cooling sequence for molecules being more aggressive than the optimal one for atoms to the fact that the molecular evaporation needs to be performed quickly to reduce the heating from technical noise and the number loss from inelastic collisions.

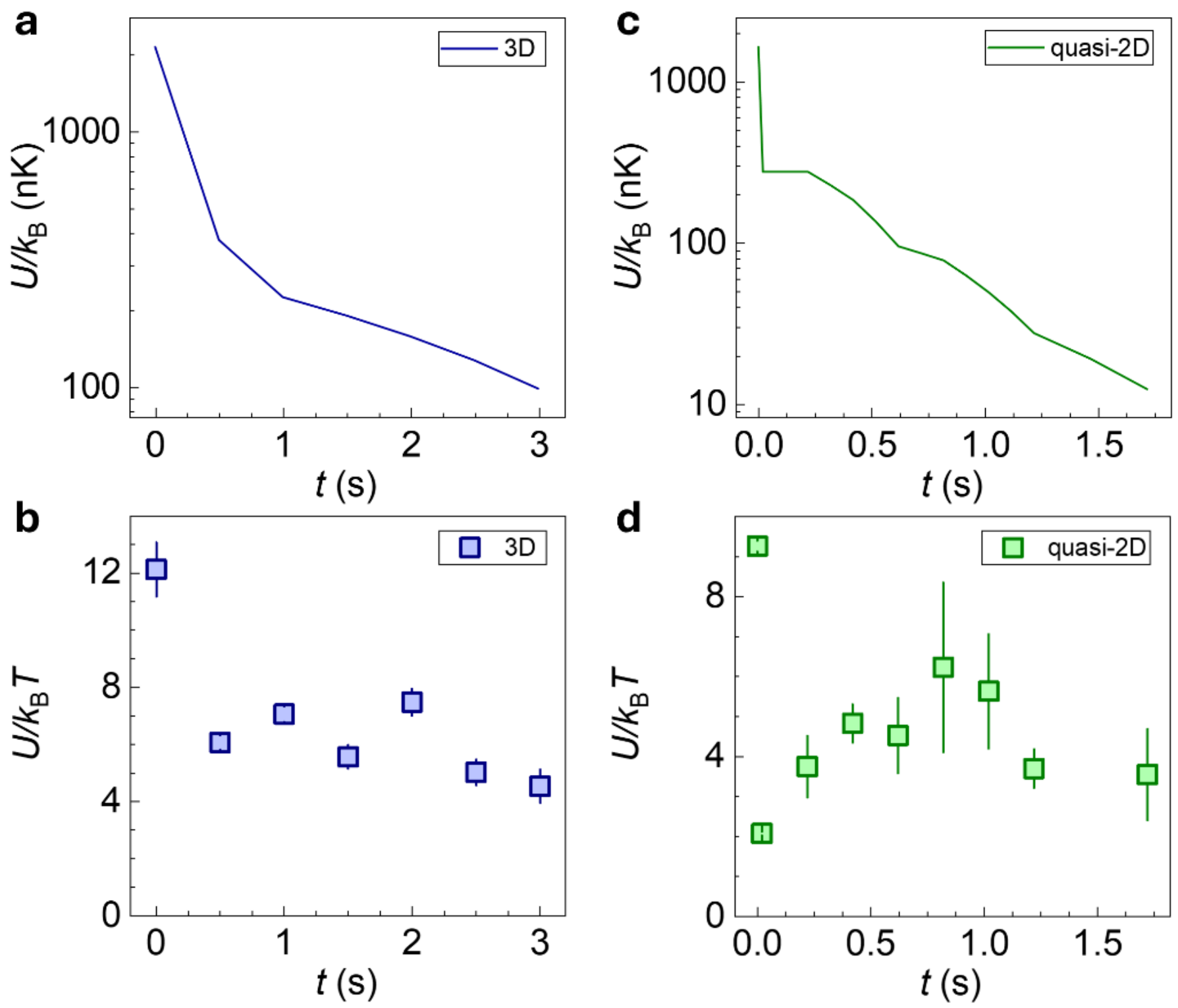

**Supplementary Fig. 7: The evaporative cooling sequence for KRb molecules in 3D and quasi-2D.** (**a**) Trap depth versus evaporation time calculated from the xODT laser power. (**b**) $U/k_\mathrm{B}T$ at different stages of the evaporation in 3D. Error bars are propagated from 1 SE of mean temperature at each step. (**c**) Trap depth of the combined electro-optical potential versus evaporation time calculated from the applied electric field curvature and VL trap potential in quasi-2D. (**d**) $U/k_\mathrm{B}T$ at different stages of the evaporation in quasi-2D. Error bars are propagated from 1 SE of mean

temperature at each step. For an error bar smaller than the size of the marker, the error (1 SE) is represented by half of the gap between two opposing sticks inside the marker.

For evaporative cooling in quasi-2D, we applied electric field curvature along the x-direction to lower the total trap depth $U$. We show $U$ as a function of the evaporation time for the optimized sequence in Supplementary Fig. 7c. After the evaporation, the electric field curvature was ramped back to zero and the temperature of the molecules was measured. Assuming that the ramping of electric field is adiabatic, the ratio $\frac{U}{k_B T}$ is extracted as shown in Supplementary Fig. 7d. $\frac{U}{k_B T}$ is around 4 throughout the evaporation in quasi-2D, again signaling that an aggressive evaporation sequence is required to minimize loss due to inelastic collisions. $\frac{U}{k_B T} \approx 4$ is in a good agreement with former measurements[1].

### J. Close-coupling scattering calculations

To theoretically obtain elastic and inelastic scattering cross sections, we utilize numerical log derivative propagation with an adaptive step size version of Johnson's algorithm[23] to solve close-coupling equations of motion[24]. We perform scattering calculations between Förster resonant shielded molecules in three-dimensions (3D) with a basis set of up to $N = 4$ rotational states and $L = 9$ partial waves to obtain numerical convergence. These 3D calculations are by now rather standard so we refer the reader elsewhere for further details[25–27]. Our calculations for two-dimensional (2D) scattering are, however, non-standard so we provide further details here.

For molecules confined in quasi-2D, a large electric field polarizes the dipoles along $z$, for which we consider their collisions when they are prepared in their field-dressed ground state $|\tilde{N}, M\rangle = |\tilde{0}, 0\rangle$. The effective interaction potential between two molecules is then:

$$V_{2\mathrm{D}}(\boldsymbol{r}) \approx -\frac{C_6}{(\rho^2 + z^2)^3} + \frac{d_{\tilde{0}}^2}{4\pi\epsilon_0} \frac{1 - 3z^2(\rho^2 + z^2)^{-1}}{(\rho^2 + z^2)^{3/2}}, \tag{19a}$$

$$d_{\tilde{N}} = d_0 \sum_{N,N'} \langle \tilde{N}, 0 | N, 0 \rangle \langle N', 0 | \tilde{N}, 0 \rangle \sqrt{(2N' + 1)(2N + 1)} \begin{pmatrix} N' & 1 & N \\ 0 & 0 & 0 \end{pmatrix}^2 \tag{19b}$$

where $\rho^2 = x^2 + y^2$ and $C_6$ is the electronic dispersion coefficient [28]. For scattering in quasi-2D, it is natural to adopt the basis

$$|\rho; n, m\rangle = \frac{u(\rho)}{\sqrt{\rho}} \frac{e^{im\phi}}{\sqrt{2\pi}} \varphi_n(z), \tag{20}$$

with $z$ harmonic oscillator states $\varphi_n(z)$. However, the relatively small dipole moment of KRb results in the basis above to be inefficient at describing the molecules at close distances[29,30].

Instead, we utilize an adiabatic coupling scheme to formulate the scattering problem in quasi-2D. Starting with the Schrödinger equation:

$$\left(-\frac{\hbar^2}{2m_r}\boldsymbol{\nabla}^2 + U(r,z)\right)\Psi(r,z,\phi) = E\Psi(r,z,\phi), \tag{21a}$$

$$U(r,z) = V_{\mathrm{2D}}(r,z) + V_{\mathrm{trap}}(z), \tag{21b}$$

we first express the Laplacian in cylindrical coordinates:

$$\boldsymbol{\nabla}^2 = \frac{1}{r}\frac{\partial}{\partial r}\left(r\frac{\partial}{\partial r}\right) + \frac{1}{r^2}\frac{\partial^2}{\partial\phi^2} + \frac{\partial^2}{\partial z^2},$$

and expand the wavefunction solutions in a quasi-adiabatic basis:

$$\Psi(r,z,\phi) = \sum_{n_z}\sum_{m_\phi}\frac{\chi\left(r;\{n_z,m_\phi\}\right)}{\sqrt{r}}\frac{e^{im_\phi\phi}}{\sqrt{2\pi}}\Phi_{n_z}(z;r), \tag{22}$$

where $\Phi_{n_z}(z;r)$are the quasi-adiabatic eigenstates that solve the equation:

$$\left(-\frac{\hbar^2}{2m_r}\frac{\partial^2}{\partial z^2} + U(r,z)\right)\Phi_{n_z}(z;r) = W_{n_z}(r)\Phi_{n_z}(z;r), \tag{23}$$

rendering the Schrödinger equation:

$$\left(-\frac{\hbar^2}{2m_r}\frac{\partial^2}{\partial r^2} + W_{n_z}(r) + \frac{\hbar^2\left(m_\phi^2 - 1/4\right)}{2m_r r^2}\right)\Phi_{n_z}(z;r)\chi(r) = E\Phi_{n_z}(z;r)\chi(r). \tag{24}$$

Cylindrical symmetry mandates that $m_\phi$ is conserved, so we do not have to bother with coupling various $m_\phi$ sectors. We utilize a discrete variable representation[31] in $z$ to obtain the adiabatic eigenstates and eigenenergies at each $r$.

Because the second derivative with respect to $r$ acts on both $\Phi_{n_z}(z;r)$ and $\chi(r)$, we have

$$\begin{aligned} E\Phi_{n_z}(z;r)\chi(r) &= -\frac{\hbar^2}{2m_r}\frac{\partial^2}{\partial r^2}\Phi_{n_z}(z;r)\chi(r) + \left(W_{n_z}(r) + \frac{\hbar^2\left(m_\phi^2 - 1/4\right)}{2m_r r^2}\right)\Phi_{n_z}(z;r)\chi(r) \\ &= -\frac{\hbar^2}{2m_r}\left(\Phi_{n_z}(z;r)\frac{\partial^2}{\partial r^2} + \frac{\partial^2\Phi_{n_z}(z;r)}{\partial r^2} + 2\frac{\partial\Phi_{n_z}(z;r)}{\partial r}\frac{\partial}{\partial r}\right)\chi(r) \\ &\quad + \left(W_{n_z}(r) + \frac{\hbar^2\left(m_\phi^2 - 1/4\right)}{2m_r r^2}\right)\Phi_{n_z}(z;r)\chi(r). \end{aligned} \tag{25}$$

for which taking the $z$ inner product of the equation above against $\Phi^*_{n'_z}(z;r)$, we obtain the radial Schrödinger equation:

$$\delta_{n'_z,n_z} E\chi(r) = \delta_{n'_z,n_z}\left(-\frac{\hbar^2}{2m_r}\frac{\partial^2}{\partial r^2} + W_{n_z}(r) + \frac{\hbar^2\left(m_\phi^2 - 1/4\right)}{2m_r r^2}\right)\chi(r)$$
$$-\frac{\hbar^2}{2m_r}\int d\,z\Phi^*_{n'_z}(z;r)\left(\frac{\partial^2 \Phi_{n_z}(z;r)}{\partial r^2} + 2\frac{\partial \Phi_{n_z}(z;r)}{\partial r}\frac{\partial}{\partial r}\right)\chi(r). \quad (26)$$

Cast in a more concise form, Eq. 26 above is instead more conveniently written as:

$$\delta_{n'_z,n_z}\left[\frac{\partial^2}{\partial r^2} + \mathcal{K}^2_{m_\phi}(r)\right]\chi(r) = \left[P_{n'_z,n_z}(r)\frac{\partial}{\partial r} + Q_{n'_z,n_z}(r)\right]\chi(r) \quad (27)$$

where

$$P_{n'_z,n_z}(r) = -2\int d\,z\left[\Phi^*_{n'_z}(z;r)\frac{\partial \Phi_{n_z}(z;r)}{\partial r}\right], \quad (28a)$$

$$Q_{n'_z,n_z}(r) = -\int d\,z\left[\Phi^*_{n'_z}(z;r)\frac{\partial^2 \Phi_{n_z}(z;r)}{\partial r^2}\right], \quad (28b)$$

$$\mathcal{K}^2_{m_\phi}(r) = \frac{2m_r}{\hbar^2}\left[E - W_{n_z}(r)\right] - \frac{m_\phi^2 - 1/4}{r^2}. \quad (28c)$$

The matrices $\boldsymbol{P}(r)$ and $\boldsymbol{Q}(r)$ above are in general impractical to compute, so instead, we obtain numerical solutions to the equation above by employing a diabatic-by-sector approach[32] to propagate the log-derivative:

$$\boldsymbol{Z}(r) = \left[\frac{\partial}{\partial r}\boldsymbol{\chi}(r)\right][\boldsymbol{\chi}^{-1}(r)], \quad (29)$$

given in terms of the fundamental solution matrix $\boldsymbol{\chi}(r)$. The radial Schrödinger equation is written in terms of $\boldsymbol{Z}(r)$ as:

$$\boldsymbol{Z}'(r) + \boldsymbol{Z}^2(r) - \boldsymbol{P}(r)\boldsymbol{Z}(r) + \left[\mathcal{K}^2_{m_\phi}(r) - \boldsymbol{Q}(r)\right] = \boldsymbol{0}, \quad (30)$$

which permit solutions via numerical log-derivative propagation. The diabatic-by-sector method of solution then proceeds as follows. The radial domain is first discretized into a set of grid points $\{r_j\}$ with $j = 1,2,3,\ldots$, whereby the potential $U(r)$ is assumed constant within each local sector $\left(r_{j-1} + r_j\right)/2 \le r \le \left(r_j + r_{j+1}\right)/2$. Within these local sectors, the effective equation of motion is then the usual diabatic one

$$\boldsymbol{Z}'(r) + \boldsymbol{Z}^2(r) + \mathcal{K}^2_{m_\phi}(r) \approx \boldsymbol{0}, \quad (31)$$

whereas the non-adiabatic couplings are included by matching the log-derivative of one sector to its previous sector at their common boundary with the transfer (i.e. passage) matrix:

$$G_{n'_z,n_z}(r_{j+1},r_j) = \int d\,z \Phi_{n'_z}(z;r_{j+1})\Phi_{n_z}(z;r_j), \tag{32}$$

such that

$$\boldsymbol{Z}^{(j+1)}(r)\big|_{r=(r_j+r_{j+1})/2} = \boldsymbol{G}(r_{j+1},r_j)\boldsymbol{Z}^{(j)}(r)\big|_{r=(r_j+r_{j+1})/2}\boldsymbol{G}^{-1}(r_{j+1},r_j), \tag{33}$$

where $\boldsymbol{Z}^{(j)}(r)$ is the log-derivative matrix in the $j$-th sector. At the first sector, we utilize absorbing boundary conditions to model loss at short range[26,33].

In order to compute the adiabatic eigenstates $\Phi_{n_z}$, the vertical direction is confined to a finite box and discretized. However, this means that adiabatic eigenstates at very high energies in the spectrum are not physical because of the hard wall boundaries of the box, so it is preferable to not use such states. For our calculation, only the lowest 200 adiabatic eigenstates (out of 16,000) within each sector are used to propagate the log derivative. If not all adiabatic eigenstates are kept however, the passage matrix is no longer necessarily unitary due to truncation of the complete basis. To deal with this issue, grid points $\{r_j\}$ are chosen adaptively such that the energy eigenvalues between sectors do not change by more than $3\%$, so that $\boldsymbol{G}(r_{j+1},r_j)$ is well approximated as unitary. We converge the calculation using $m_\Phi = 1,3,5,7,9$. The results show good agreement with Ref. 30.

### K. Evaporative cooling simulations

We simulate the evaporative cooling of KRb molecules in 2D and 3D using a direct simulation Monte Carlo approach as was implemented in Ref. 34. Put briefly, the molecular distribution is discretized into simulated particles that are taken to undergo motion in classical phase space, progressing forward in discrete time steps of $\Delta t$ via Störmer-Verlet symplectic integration[35] in the presence of the background optical dipole trap. The ODT potential energy surface is modeled from the Gaussian beams used in the experiment[36], with gravity included in 3D and an antiharmonic potential from electric field curvature included in 2D. The trap parameters vary over time to lower the trap depth and force evaporation, just as they are in the experiment.

Molecular collisions are simulated by Monte Carlo sampling of collision events, based on the scattering cross sections obtained from numerical close-coupling calculations (see Sec. Close-coupling scattering calculations). Molecular losses from inelastic collisions, evaporation, and one-body decay are included by removing simulated particles from the numerical particle ensemble during the evaporation sequence. Fermi statistics is encoded into the simulated collisions as described in Refs. [34,37–39], where sampled collisions are retroactively blocked based on the local phase space occupancy of where collided particles would have scattered into. Once blocked, the collision event is undone, and the particles progress as though having never collided.

We also perform evaporation simulations that compensate for the effect of Pauli suppression on thermalization, allowing estimates of its consequence on the evaporation efficiency. In these artificially modified simulations, the elastic cross section is dynamically enlarged through the evaporation trajectory by multiplication with $\eta^{-1}$ as a function of $T/T_{\mathrm{F}}$. Applying this factor then compensates for the effect of Pauli blocking on the thermalization rate, while maintaining Fermi-Dirac statistics on the molecular distribution for thermometric consistency with the physically accurate non-compensated simulation. By comparing the compensated and non-compensated simulations under the same evaporation sequence, we isolate the effect of a filled Fermi sea on the evaporative cooling efficiency due to suppressed thermalization.

Lastly, we note that the simulations in 2D are purely 2D in the sense that molecules only have two position coordinates. This assumption is sufficient because temperatures are low enough that almost all molecules occupy the ground band of the tightly confining vertical direction. Furthermore, the cross sections used to sample collisions are calculated for an actual quasi-2D system with harmonic confinement in the vertical direction. For the bilayer experimental sample, the simulation therefore only tracks one of the layers for the evaporation and ignores the interaction between layers, as the interlayer elastic scattering cross section is estimated to be at least an order of magnitude lower than the intralayer cross section at a temperature of 10 nK.

### L. Fermi suppression of thermalization

As the phase space density of a Fermi gas increases, the rate of collisional thermalization becomes suppressed due to the diminishing phase space volume that molecules can scatter into. More collisions are thus required for populating the available phase space regions for the gas to reach equilibrium. We can quantify this reduction in thermalization by a Fermi suppression factor $0 \leq \eta \leq 1$, that compares the collision rate expected in a classical gas, to that in a Fermi degenerate gas[40]:

$$\gamma = \eta \frac{n\langle \sigma v_r \rangle}{\mathrm{N_{col}}}, \tag{34}$$

where the standard collision rate $n\langle \sigma v_r \rangle$ and the number of collisions per rethermalization $\mathrm{N_{col}}$[41] are both defined assuming Maxwell-Boltzmann statistics. In general, $\mathrm{N_{col}}$ is a tensor with indices corresponding to the axes of excitation and rethermalization[18]. However, we will only consider a single geometry excitation-rethermalization set by the experiment here, allowing us to ignore the tensor indices of $N_{col}$. Above, $n$ is the mean density, $\sigma$ is the integral elastic cross section and $v_r$ is the relative collision velocity. To compute $\eta$, we utilize a variational approach[40,42] by first computing the rethermalization rate:

$$\gamma \approx -\frac{\langle \mathcal{J}[\Phi]\Phi \rangle_F}{2\langle \Phi^2 \rangle_F}, \tag{35}$$

where $\langle \dots \rangle_F = \int d^{\mathrm{d}}\boldsymbol{r} d^{\mathrm{d}}\boldsymbol{p} f(\boldsymbol{r},\boldsymbol{p})\left[1 - h^{\mathrm{d}} f(\boldsymbol{r},\boldsymbol{p})\right](\dots)$, d is the number of coordinate dimensions, and $\Phi = \Phi(\boldsymbol{p})$ is a variational quantity. Above, $\mathcal{J}[\Phi] = \mathcal{C}[\Phi]/\left[f\left(1 - h^{\mathrm{d}} f\right)\right]$ is defined in terms of the collision integral[43]

$$\mathcal{C}[\Phi] = \int \frac{d^{\mathrm{d}}\boldsymbol{p}_1}{m} |\boldsymbol{p} - \boldsymbol{p}_1| f(\boldsymbol{r},\boldsymbol{p}) f(\boldsymbol{r},\boldsymbol{p}_1) \int d\Omega' \frac{d\sigma_{\mathrm{el}}}{d\Omega'} \left(1 - h^{\mathrm{d}} f'\right)\left(1 - h^{\mathrm{d}} f'_1\right) \Delta\Phi, \tag{36}$$

where $d\sigma_{\mathrm{el}}/d\Omega'$ is the elastic differential cross section with the d-dimensional solid angle of post-collision relative momentum coordinates $d\Omega'$. We utilize $\Phi = p_i^2 - \boldsymbol{p}^2/\mathrm{d}$ as the variational function, where the subscript $i$ denotes the axis of initial gas excitation.

**M. Thermalization considerations for evaporative cooling**

Evaporative cooling is expected to become inefficient when the thermalization rate (Eq. 34) becomes smaller than the rate at which the trap depth is lowered during the evaporation sequence. At this point, forced evaporation happens faster than the gas can thermalize, which has two consequences. First, the thermal wings of the distribution are not repopulated during forced evaporation, so the cooling efficiency decreases due to having fewer "hot" molecules available to evaporate. Secondly, evaporation in both 2D and 3D proceeds along a single axis, so forced evaporation at a rate faster than the thermalization rate can affect the dimensionality of evaporation[22] and also reduce its cooling efficiency. In theory, lowering the trap depth infinitely slowly would remove these problems, but in practice, the actual sequence time is limited to second scale by two-body loss and technical heating. As a result, the $T/T_{\mathrm{F}}$ is expected to saturate during the sequence, once the evaporation becomes inefficient.

Examining Eq. 34, the decrease in the thermalization rate can be attributed to a decrease in $\eta n\langle \sigma v_r \rangle$, or to an increase in $N_{\mathrm{col}}$. The value of $N_{\mathrm{col}}$ is constant in the threshold scattering regime, assuming a Maxwell-Boltzmann ensemble of molecules.

Furthermore, as can be seen in Fig. 2d and Fig. 3d, the value for $\sigma$ is fairly constant in both quasi-2D and 3D throughout the evaporation trajectory. The number density $n$ is approximately proportional to the number of molecules but is inversely related to the temperature, resulting in $n$ remaining fairly constant throughout the evaporation in both cases. Therefore, it seems that the main difference in quasi-2D and 3D is the Fermi suppression factor $\eta$ and $\langle v_r \rangle$. In the classical limit, the scaling of $\langle v_r \rangle$ is proportional to $\sqrt{T}$, and in the limit of deep degeneracy, this velocity is roughly set by the Fermi velocity $v_F \propto \sqrt{E_\mathrm{F}}$, which scales as $N^{1/4}$ in 2D and $N^{1/6}$ in 3D. Finally, the Fermi suppression factor $\eta$ can be seen in Fig. 1e. Overall, these temperature and number scalings cause the thermalization rate to decrease by two to three orders of magnitude during forced evaporation.

From this discussion, it appears that the variation of the thermalization rate over an evaporation trajectory is not drastically different in quasi-2D versus 3D. The reason that the evaporation efficiency plateaus more obviously in quasi-2D than in 3D is due to the absolute value of the thermalization rate, which is much lower in 3D due to the smaller dipole moment of the $|1,0\rangle$ rotational state at the Förster resonant shielding point, compared to the ground state dipole moment at 6.5 kV/cm used in quasi-2D. At the beginning of the evaporation sequence, the value of $n\langle \sigma v_r \rangle$ is approximately 10 $s^{-1}$ for the 3D initial conditions and 1000 $s^{-1}$ for the quasi-2D initial conditions. These rates differ by around two orders of magnitude, whereas the rate at which the sequence proceeds in quasi-2D versus 3D is similar, set by technical heating rates of around 2 nK/s in both cases. As a result, the more favorable elastic to inelastic ratio in quasi-2D makes the technical heating relatively less important, causing a plateau of the evaporation efficiency at a lower $T/T_\mathrm{F}$ in quasi-2D compared to 3D.

## Supplementary References